\documentclass[a4paper,11pt]{article}
\pdfoutput=1 % if your are submitting a pdflatex (i.e. if you have
\usepackage{jcappub} % for details on the use of the package, please
\usepackage[T1]{fontenc} % if needed

\title{Measuring Line-of-sight Distances to Haloes with Astrometric Lensing B-mode }

\usepackage{amsmath,amssymb}
\usepackage{graphicx}
\usepackage{units}
\usepackage{color}
\usepackage{bm} 
\usepackage{subcaption}

\newcommand{\del}{\partial}

\newcommand{\ALP}{{\boldsymbol{\alpha}}}

\newcommand{\x}{{\boldsymbol{x}}}
\newcommand{\y}{{\boldsymbol{y}}}

\newcommand{\R}{{\boldsymbol{r}}}

\newcommand{\BF}{\begin{figure}\begin{center}}
\newcommand{\EF}{\end{center}\end{figure}}
\newcommand{\BE}{\begin{equation}}
\newcommand{\EE}{\end{equation}}
\newcommand{\BEA}{\begin{eqnarray}}
\newcommand{\EEA}{\end{eqnarray}}

\def\v#1{\boldsymbol #1}
\newcommand{\bvec}[1]{\mbox{\boldmath $#1$}}
\newcommand{\ms}{M_{\odot}}

\usepackage{etoolbox}

 \author{Kaiki Taro Inoue}
\affiliation{Faculty of Science and Engineering, Kindai University,\\Higashi-Osaka, 577-8502, Japan}

\emailAdd{kinoue@phys.kindai.ac.jp}
\abstract{Relative astrometric shifts between multiply lensed images provide a valuable tool to investigate haloes in the intergalactic space. In strong lens systems in which a single lens plays the primary role in producing multiple images, the gravitational force exerted by line-of-sight (LOS) haloes can slightly change the relative positions of multiply lensed images produced by the dominant lens. In such cases, a LOS halo positioned sufficiently far from the dominant lens along the LOS can create a pattern in the scaled deflection angle that corresponds to the B-mode (magnetic or divergence-free mode). By measuring both the B-mode and E-mode (electric or rotation-free mode), we can determine the LOS distance ratios, as well as the 'bare' convergence and shear perturbations in the absence of the dominant lens. However, scale variations in the distance ratio lead to mass-sheet transformations in the background lens plane, introducing some uncertainty in the distance ratio estimation. This uncertainty can be significantly reduced by measuring the time delays between the lensed images. Additionally, if we obtain the redshift values of both the dominant and perturbing haloes, along with the time delays between the multiply lensed images that are affected by the haloes, the B-mode can break the degeneracy related to mass-sheet transformations in both the foreground and background lens planes. Therefore, measuring the astrometric lensing B-mode has the potential to substantially decrease the uncertainty in determining the Hubble constant.}

\begin{document}
\maketitle
\flushbottom

\section{Introduction}
Cold dark matter (CDM) models encounter challenges on scales below $\sim 100\,$kpc, particularly concerning dwarf galaxies. For instance, the observed number count of dwarf satellite galaxies in nearby galaxies in the Local Group falls significantly short of the predicted number of subhaloes capable of hosting dwarf galaxies as projected by CDM models \cite{kauffmann1993, klypin1999, moore1999}. Hydrodynamic simulations that incorporate baryonic feedback and cosmic reionization have emerged as potential solutions to this discrepancy for the Milky Way \cite{wetzel2016, brooks2017, fielder2019}. However, it remains uncertain whether the Milky Way represents a "typical galaxy," and the extent to which this discrepancy can be explained for other galaxies remains an open question \cite{nashimoto2022}. It is plausible that dark matter differs from CDM, and there may be fewer low-mass haloes on scales below $\sim 100\,$kpc capable of hosting dwarf galaxies than CDM predictions suggest.

Gravitational lensing serves as a potent tool for investigating low-mass haloes with masses $\lesssim 10^9\,\ms$ in the distant universe. In particular, the study of quadruply lensed quasar-galaxy and galaxy-galaxy strong lens systems has proven valuable for probing low-mass haloes. This is because the strong lensing effect induced by a foreground galaxy amplifies the weak lensing signals of low-mass haloes, which are otherwise challenging to detect without such enhancement. Typically, the relative positions of lensed images can be accurately fitted using a smooth gravitational potential within a few milliarcseconds. However, in some of these systems, the flux ratios of the lensed images in radio or mid-infrared wavelengths deviate by 10-40\% from the predictions of the theoretical models. It has been suggested that such anomalies in flux ratios, particularly in radio or mid-infrared emissions, may be attributed to subhaloes residing within the host galaxy \cite{mao1998, metcalf2001, chiba2002, dalal-kochanek2002, keeton2003, inoue-chiba2003, inoue2005a, xu2009, xu2010}. Nevertheless, it is worth noting that these anomalies might also find explanations in the complex gravitational potential of the foreground galaxy \cite{evans2003, oguri2005, gilman2017, hsueh2017, hsueh2018}. The interpretation gains support from discrepancies observed in the relative astrometric shifts of lensed extended images \cite{treu-koopmans2004, koopmans2005, vegetti2009, vegetti2010, chantry2010, vegetti2012, vegetti2014, inoue-minezaki2016, hezaveh2016a, chatterjee2018, cagan2020}.

However, any small-mass haloes along the line of sight (LOS) can also influence the flux ratios and relative positions of lensed images \cite{metcalf2005a, xu2012}. Based on a semi-analytical approach, \cite{inoue-takahashi2012} argued that the primary cause of anomalies in flux ratios for quadruply lensed quasars with an Einstein radius of approximately $1\,$arcsec is the presence of small-mass haloes within the intergalactic space, rather than subhaloes within the foreground galaxy. This assertion was subsequently confirmed by \cite{takahashi-inoue2014}, which employed $N$-body simulations capable of resolving dwarf galaxy-sized haloes within a cosmological context. Building on semi-analytical approach, \cite{inoue2016} highlighted that for source redshifts $z_\textrm{s}>1$, the cumulative effect of line-of-sight structures, including haloes and troughs, on altering the flux ratios of quadruply lensed images ranges from 60 to 80 percent in the CDM models. A subsequent analysis arrived at a similar conclusion \cite{despali2018}.

The observational evidence supporting the significant role of LOS haloes in 'substructure-lensing' is mounting. Based on $N$-body simulations, \cite{takahashi-inoue2014} pointed out that the 'object X', previously assumed to be a satellite galaxy in the lensed quasar MG\,J0414+0534, may in fact be located within the intergalactic space. Additionally, \cite{sengul2022} argued that a presumed dark perturber, assumed to be a satellite galaxy in the lensed quasar B1938+666, actually constitutes an intergalactic halo with a mass of approximately $2\times 10^9\,\ms$. Furthermore, a recent assessment of the lensing power spectra at an angular wave number of $l=1.2\times 10^6$ or roughly $9\,$kpc within the primary lens plane aligns with CDM models, where LOS haloes exert the predominant influence on the alterations in flux and relative positions of lensed images \cite{inoue2023}.

To validate this claim, it becomes imperative to determine line-of-sight (LOS) distances to perturbing dark haloes. This necessitates the measurement of astrometric shifts between perturbed multiple images and their unperturbed counterparts. These astrometric shifts have the potential to break the degeneracy between subhalo mass and the LOS distance, provided that they are resolved at the scale of the Einstein radius of the perturbing object. Additionally, precise determination of sky positions using a dipole structure in the intensity of the lensed extended image is required \cite{inoue2005b, inoue2013}.

In practical scenarios, the presence of observable dipole structures or a fifth image due to the strong lensing effects caused by intergalactic small-mass haloes is rare. Moreover, numerous perturbers with varying masses, residing at different redshifts and sky positions, can influence astrometric shifts. Consequently, modeling each dark perturber individually becomes a formidable challenge. To achieve precise modeling of astrometric shifts, we must comprehend the coupling effect between the strong lensing exerted by a dominant primary lens and the weak lensing introduced by a subdominant secondary lens located \textit{outside} the primary lens plane. While frameworks for modeling 'LOS lensing' by intergalactic haloes have been explored in previous literature \cite{erdl1993, rennan1996, mccully2014, mccully2017, birrer2017, fleury2021}, none of these analyses have fully addressed the coupling effect for general perturbations, encompassing all multipole moments (beyond flexions). Particularly, the ambiguity of LOS distances stemming from multi-plane mass-sheet transformations \cite{schneider2014, schneider2019} remains poorly understood.

In this paper, we investigate the property of B-modes (magnetic modes) in the two-dimensional vector field of astrometric shifts as a means to determine the distance ratio to a perturber. The concept is straightforward: if all perturbers are confined to the primary lens plane, their reduced deflection angles\footnote{The reduced deflection angle is defined as the angle between the true position of a source and the observed position of the lensed image.} can be expressed as gradients of a scalar potential, resulting in no generation of B-modes. However, if some perturbers are residing at a different lens plane, the reduced deflection angles of those perturbers that \textit{are assumed to be} in the primary lens plane cannot be represented as gradients of a scalar potential due to the coupling effect. Consequently, this generates B-modes akin to those induced by weak lensing in the cosmic microwave background (CMB) polarization \cite{zaldarriaga1998, lewis2006}. It should be noted that, in our scenario, perturbers can exist in the background of the primary lens, whereas in weak lensing of the CMB, perturbers can only reside in the foreground of the CMB. In Section 2, we present the theoretical framework of our method. In Section 3, we explore how multi-plane mass-sheet transformations impact the LOS distance ratio and time delay in a double lens system. In Section 4, we examine the astrometric Lensing B-mode and evaluate the accuracy of distance ratio estimators using simple toy models. Finally, in Section 5, we offer conclusions and discuss the observational feasibility of our proposed method.

%\begin{figure}
%
%\includegraphics[width=8.3cm,pagebox=cropbox,clip]{f1.pdf}
%\vspace{-0.8cm}
%\caption{Schematic picture of haloes (yellow disks) and troughs (blue bars) in %sight lines. The cylinder represents bundles of light rays that pass the vicini%ty of an Einstein ring in the lens plane.  }
%\label{fig:trough.pdf}
%\end{figure} 

\section{Theory}
\subsection{E/B decomposition}
Helmholtz' theorem states that a smooth three dimensional vector field $\ALP(\R), \R=(x_1,x_2,x_3)$ that decays faster than $|\R|^{-2}$ for $|\R|\rightarrow \infty$ can be uniquely decomposed as a sum of a rotation free 'electric' field $\ALP^\textrm{E}$ and a divergence free 'magnetic' field $\ALP^\textrm{B}$
as $\ALP=\ALP^\textrm{E}+\ALP^\textrm{B}$. By constraining $\alpha_3$ to be a constant, we can apply the theorem to two dimensional vectors: In terms of an 'electric' potential $\psi^\textrm{E}$ and a 'magnetic' potential $\psi^\textrm{B}$, a smooth two dimensional vector field $\ALP(\x), \x=(x_1,x_2)$ can be decomposed as
\BEA
\alpha_i&=&\psi^\textrm{E}_{,i}+\varepsilon_i^j \psi^\textrm{B}_{,j}, 
\\
&=&\alpha^{\textrm{E}}_i+\alpha^{\textrm{B}}_i,  ~~(i,j=1,2),
\EEA 
where $\varepsilon$ is the Levi-Civita tensor and $,i$ is a partial derivative operator in the direction $x_i$. For brevity, we denote the two dimensional rotation as $\nabla \times \ALP^\textrm{B}\equiv \alpha^{\textrm{B}}_{2,1}-\alpha^{\textrm{B}}_{1,2}=-\psi^\textrm{B}_{,1,1}-\psi^\textrm{B}_{,2,2}$ and divergence as $\nabla \cdot \ALP^\textrm{E}\equiv \alpha^{\textrm{E}}_{1,1}+\alpha^{\textrm{B}}_{2,2}=\psi^\textrm{E}_{,1,1}+\psi^\textrm{E}_{,2,2}$. 

In the following, we consider a lens system in which a source at a redshift $z_\textrm{s}$ is lensed by a primary (dominant) lens at a redshift $z_\textrm{d}$ with a reduced deflection angle $\bvec{\alpha}$ and a secondary lens (perturber) at a redshift $z_\textrm{f}(<z_\textrm{d})$ or $z_\textrm{b}(> z_\textrm{d})$ with a small reduced deflection angle $\delta \bvec{a}\,(|\delta \bvec{\alpha}| \ll |\bvec{\alpha}| )$. If the perturber resides at the lens plane of the dominant lens, i.e., $z_\textrm{P}=z_\textrm{d}$, $\nabla \times \delta \bvec{\alpha}$ is null. However, if the 
perturber does not reside at the lens plane, the rotation of the effective deflection angle\footnote{The effective deflection angle is defined as the astrometric shift due to a secondary lens with respect to the position of the de-lensed image in a primary lens.} $\nabla \times \delta \bvec{\alpha}_\textrm{eff}$ is not null. We denote the angular diameter distance between a dominant lens and the source as $D_{\textrm{sd}}$, between a foreground perturber and a dominant lens as $D_{\textrm{df}} (z_\textrm{f}<z_\textrm{d})$,  between a dominant lens and a background perturber as $D_{\textrm{bd}}(z_\textrm{b}>z_\textrm{d})$, between a foreground perturber and the source as $D_{\textrm{sf}}$, between a background perturber and the source as $D_{\textrm{sb}}$.

\subsection{Rotation and divergence}
First, we consider a double lens system in which a perturber resides in front of a dominant lens, i.e., $z_\textrm{f}<z_\textrm{d}$.  The lens equation for the dominant lens with a deflection angle $\v{\alpha}$ is
\BE
\y'=\x-\v{\alpha}(\x), \label{eq:lensd}
\EE
where $\x$ and $\y'$ are the angular coordinates in the dominant lens plane and the source plane, respectively. If the deflection angle is perturbed by a foreground perturber
by $\delta \bvec{\alpha}$, then the perturbed angular position $\bvec{y}$ can be written as
\BE
\y=\x-\delta \bvec{\alpha}(\bvec{x})-\bvec{\alpha}(\y_\textrm{d}), \label{eq:lenseq-f}
\EE
where $\y_\textrm{d} \equiv \bvec{x}-\beta\, \delta \bvec{\alpha}(\bvec{x}), (0<\beta<1) $ and
\BE
\beta=\frac{D_\textrm{df} D_\textrm{s}}{D_\textrm{d} D_\textrm{sf} },
\EE
is the LOS distance ratio parameter, which encodes the information of the LOS distance from an observer to the perturber in the foreground with respect to the dominant lens. If we interprete the system as the one in which a perturber with an effective deflection angle $\delta \bvec{\alpha}_{\textrm{eff}}^{\textrm{f}}$ resides at the dominant lens plane, eq.(\ref{eq:lenseq-f}) can be written as
\BEA
\bvec{y}&=&\bvec{x}-\ALP(\x)-\delta \ALP_{\textrm{eff}}^{\textrm{f}}(\x),
\nonumber
\\
\delta \ALP_{\textrm{eff}}^{\textrm{f}}(\x)&\equiv &\delta \ALP(\x)+\ALP(\y_\textrm{d}(\x))-\ALP(\bvec{\x}).
 \label{eq:lenseq-f2}
\EEA
Eqs. (\ref{eq:lensd}) and (\ref{eq:lenseq-f2}) gives an astrometric shift $\delta \y$ at $\x$,  
\BEA
\delta \y &=&\y-\y' \nonumber
\\
&=&-\delta \ALP_{\textrm{eff}}^{\textrm{f}}(\x).
\EEA
We note that $\delta \y$ is a function of $\x$\footnote{If $\delta \y$ is a function of $\y$, not $\x$, the astrometric shifts are not observable. Such astrometric shifts should strongly correlate with the positions of multiply lensed images of a dominant lens: The determinant of $d \delta \y/d \x$ vanishes on the critical curves of the dominant lens as $\det(d \delta \y/d \x)= \det (d \delta \y/d \y)\det (d \y/d \x)=0$. Since LOS perturbers do not have correlation with the dominant lens, such unobservable shifts are considered to be unphysical.}, which yields a field of astrometric shift $\delta \y$ at $\y'$ in each set of de-lensed multiple images. For instance, if the dominant lens produces a quadruple image, the number of de-lensed images is four. In this case, there are four astrometric shift fields, where each one of them is defined within a copy of the diamond caustic.         

Because of coupling between the strong lensing by the dominant lens and the weak lensing by the foreground perturber, the effective deflection angle $\delta \ALP_{\textrm{eff}}^{\textrm{f}}$ has a magnetic component if and only if $\beta >0$. The rotation of $\delta \ALP_{\textrm{eff}}^{\textrm{f}}$ is 
\BE
\nabla \times \delta \ALP_{\textrm{eff}}^{\textrm{f}}(\x)=2 \beta (\tilde{\gamma}_1\delta \gamma_2-\tilde{\gamma}_2\delta \gamma_1),
\label{eq:lenseq-f2N}
\EE
where $\gamma_i$ is the $i$-component of the shear tensor of the dominant lens at $\tilde{\x}$ and $\delta \tilde{\gamma}_i$ is the $i$-component of the shear tensor of the secondary lens at $\x$.  

Second, we consider a double lens system, in which a perturber resides in the background of a dominant lens, i.e., $z_\textrm{b}>z_\textrm{d}$. Then, the angular position $\y$ of the source can be written as a function of the angular position $\x$ of the lensed image as
\BE
\y=\x-\ALP(\x)-\delta \ALP(\y_\textrm{b}), \label{eq:lenseq-b}
\EE
where $\y_\textrm{b} \equiv \x-\xi\, \ALP(\x), (0<\xi<1) $ and
\BE
\xi=\frac{D_\textrm{bd} D_\textrm{s}}{D_\textrm{b} D_\textrm{sd} },
\EE
is the LOS distance ratio parameter, which encodes the information of the LOS distance from an observer to perturber in the background of the dominant lens. 

As in the foreground case, we can model the system as the one in which a secondary perturber with an effective deflection angle $\delta \bvec{\alpha}_{\textrm{eff}}^{\textrm{b}}$ resides at the dominant lens plane, eq. (\ref{eq:lenseq-b}) can be written as
\BEA
\y&=&\x-\ALP(\x)-\delta \ALP_{\textrm{eff}}^{\textrm{b}}(\x),
\nonumber
\\
\delta \ALP_{\textrm{eff}}^{\textrm{b}}(\x)&\equiv &\delta \ALP(\y_\textrm{b}).
 \label{eq:lenseq-b2}
\EEA
Because of the difference between the angular position of the lensed image $\x$
and that of the background perturber $\y_\textrm{b}$, $\delta \ALP_{\textrm{eff}}^{\textrm{b}}(\x)$ has a magnetic component if and only if $\xi>0$. The rotation of $\delta \ALP_{\textrm{eff}}^{\textrm{b}}$ is 
\BE
\nabla \times \delta \ALP_{\textrm{eff}}^{\textrm{b}}(\x)=- 2 \xi (\gamma_1 \delta \gamma_2-\gamma_2 \delta \gamma_1),
\label{eq:lenseq-b2N}
\EE
where $\gamma_i$ is the $i$-component of the shear tensor of the dominant lens at $\x$ and $\delta \gamma_i$ is the $i$-component of the shear tensor of the dominant lens at $\y_\textrm{b}$

Ignoring second order terms, eqs. (\ref{eq:lenseq-f2N}) and (\ref{eq:lenseq-b2N}) can be combined to yield 
\BE
\eta= \frac{\nabla \times \delta \ALP_{\textrm{eff}}(\x)} {2 (\gamma_1 \delta \gamma_2-\gamma_2 \delta \gamma_1)},
\label{eq:eta}
\EE
where $\delta \ALP_{\textrm{eff}}(\x)$ is defined as $\delta \ALP_{\textrm{eff}}(\x) \equiv \delta \ALP_{\textrm{eff}}^\textrm{f}(\x)$, $\eta \equiv \beta$ for $\eta>0 $, 
and $\delta \ALP_{\textrm{eff}}(\x)\equiv \delta \ALP_{\textrm{eff}}^\textrm{b}(\x)$, $\eta \equiv -\xi$ for $\eta <0 $. The range of $\eta $ is $-1<\eta <1$.

Thus, from measured rotation of the effective deflection angle and shears of the dominant lens and the perturber, we can measure the dimensionless distance ratio parameter $\eta $, that encodes the information of LOS distance to the perturber. However, it should be noted that $\eta$ is not a directly measured quantity as $\delta \gamma_1$ and $\delta \gamma_2$ are not directly measured. Therefore, we need to estimate $\eta$ using a certain approximation, which we will discuss in the next section. If $\eta$ is negative/positive, the perturber resides in the background/foreground of the dominant lens. In shear-aligned coordinates $(x'_1,x'_2)$ in which the shear of the dominant lens is aligned with $x'_1$ or $x'_2$ axis, i.e. $\gamma_2'=0$, the 
rotation is proportional to $\delta \gamma_2'$. In other words, the amplitude of rotation is proportional to the non-diagonal shear component of the perturber.

In a similar manner, we can derive the divergence of the effective deflection $\delta \ALP_{\textrm{eff}}$. In the foreground and background cases, the divergences are
\BE
\nabla \cdot \delta \ALP_{\textrm{eff}}^{\textrm{f}}(\x)=2\delta \kappa-2 \beta (\tilde{\kappa} \delta \kappa+\tilde{\gamma_1}\delta \gamma_1+\tilde{\gamma_2}\delta \gamma_2), \label{eq:divf}
\EE
and 
\BE
\nabla \cdot \delta \ALP_{\textrm{eff}}^{\textrm{b}}(\x)=2\delta \kappa-2 \xi (\kappa \delta \kappa+\gamma_1\delta \gamma_1+\gamma_2\delta \gamma_2), \label{eq:divb}
\EE
respectively. Ignoring second order terms, in terms of $\eta$, eqs. (\ref{eq:divf}) and (\ref{eq:divb}) give a modified Poisson equation,
\BEA
\nabla \cdot \delta \ALP_{\textrm{eff}}(\x)&=&2\delta \kappa-2 |\eta| (\kappa \delta \kappa+\gamma_1\delta \gamma_1+\gamma_2\delta \gamma_2),
\nonumber
\\
&=&2\delta \kappa_{\textrm{eff}},\label{eq:poission}
\EEA
where $\delta \kappa_{\textrm{eff}}$ is the effective convergence perturbation which encodes the information of coupling between the dominant lens and the perturber. 

The effective deflection angle can be decomposed as magnetic and electric components,
\BE
\delta \ALP_{\textrm{eff}}=\delta \ALP^\textrm{B}_{\textrm{eff}}+\delta \ALP^\textrm{E}_{\textrm{eff}},
\EE 
where $\nabla \cdot \delta \ALP^\textrm{B}_{\textrm{eff}}=\nabla \times \delta \ALP^\textrm{E}_{\textrm{eff}}=0$.

Using eqs. (\ref{eq:eta}) and (\ref{eq:poission}), we can estimate the ratio of the
magnetic component to the electric one in the limit $\eta \rightarrow 0$ as 
\BEA
r^\textrm{BE}&\equiv& \frac{|\delta \ALP^\textrm{B}_{\textrm{eff}}|}{|\delta \ALP^{\textrm{E}}_\textrm{\textrm{eff}}|}
\nonumber
\\
&\sim & \frac{|\nabla \times \delta \ALP^\textrm{B}_{\textrm{eff}}|}{|\nabla \cdot \delta \ALP^\textrm{E}_\textrm{\textrm{eff}}|}
\nonumber
\\
&\sim& \frac{|\eta\, \gamma'_1|}{\sqrt{2}},
\label{eq:etaest1}
\EEA
where prime denotes the coordinates in which the magnification matrix of the dominant lens is diagonalized, i.e., $\gamma'_2=0$ and we assumed that $|\delta \gamma'_2|\sim |\delta \kappa'|/\sqrt{2}$. Let us suppose a typical lens system in which the dominant lens is modelled by a singular isothermal sphere (SIS). In the vicinity of an Einstein ring, the shear is $\gamma'_1\sim 0.5$. Then the ratio is 
\BE
r^\textrm{BE}\sim \frac{\sqrt{2}|\eta|}{4}. 
\label{eq:etaest2}
\EE
Let us suppose a lens system with a perturber at $|\eta|=0.4$. Then we have  $r^\textrm{BE} \sim 1/7$. Thus the contribution from magnetic component is expected to be subdominant unless the perturber is near the source or observer. 
\subsection{Estimators of LOS distance}
In order to estimate the LOS distance ratio $\eta$, we need to assess the components $\delta \kappa$, $\delta \gamma_1$, and $\delta \gamma_2$ using the spatial derivatives of the effective deflection $\delta \ALP_{\textrm{eff}}$,
\BE
\delta M_\textrm{eff}=\begin{pmatrix}
   A & B \\
   C & D
\end{pmatrix} \equiv 
\begin{pmatrix}
   \delta \alpha_{\textrm{eff}1,1} & \delta \alpha_{\textrm{eff}1,2} \\
   \delta \alpha_{\textrm{eff}2,1} & \delta \alpha_{\textrm{eff}2,1}
\end{pmatrix}. \label{eq:ABCD}
\EE

First, we consider the foreground perturber case. From eqs. (\ref{eq:lenseq-f2}, \ref{eq:ABCD}), neglecting second order terms, we have
\BEA
A&=& \delta \kappa+\delta \gamma_1-\beta [(\kappa+\gamma_1)(\delta \kappa+\delta \gamma_1)+\gamma_2 \delta \gamma_2]
\nonumber
\\
&+&\tilde{\kappa}+\tilde{\gamma_1}-\kappa-\gamma_1,
\nonumber
\\
B&=& \delta \gamma_2-\beta (\kappa\delta \gamma_2+\gamma_2\delta \kappa
+\gamma_1\delta \gamma_2-\gamma_2\delta \gamma_1)
\nonumber
\\
&+&\tilde{\gamma}_2-\gamma_2,
\nonumber
\\
C&=& \delta \gamma_2-\beta (\kappa\delta \gamma_2+\gamma_2\delta \kappa
+\gamma_2 \delta \gamma_1-\gamma_1\delta \gamma_2)
\nonumber
\\
&+&\tilde{\gamma}_2-\gamma_2,
\nonumber
\\
D&=& \delta \kappa-\delta \gamma_1-\beta [(\kappa-\gamma_1)(\delta \kappa-\delta \gamma_1)+\gamma_2 \delta \gamma_2]
\nonumber
\\
&+&\tilde{\kappa}-\tilde{\gamma_1}-\kappa+\gamma_1.
\label{eq:ABCD2}
\EEA
In first order in $\delta \ALP$, the last terms that represent small changes in the magnification matrix of 
the dominant lens can be linearly approximated as  
\BEA
c_1 &\equiv&  \frac{\del (\kappa+\gamma_1)}{\del \x} \biggr|_\x \cdot \delta \ALP(\x)
\approx -\beta^{-1}(\tilde{\kappa}+\tilde{\gamma_1}-\kappa-\gamma_1),
\nonumber
\\
c_2 &\equiv&
\frac{\del (\kappa-\gamma_1)}{\del \x} \biggr|_\x \cdot \delta \ALP(\x)
\approx -\beta^{-1}(\tilde{\kappa}-\tilde{\gamma_1}+\kappa-\gamma_1),
\nonumber
\\
c_3 &\equiv&
\frac{\del \gamma_2}{\del \x} \biggr|_\x \cdot \delta \ALP(\x)
\approx -\beta^{-1} (\tilde{\gamma_2}-\gamma_2).
\nonumber
\\
\label{eq:c1c2c3}
\EEA 
Let us suppose that a perturber resides in the vicinity of the dominant lens, i.e., $\beta \ll 1$. Then
the magnetic component of $\delta \ALP_\textrm{eff}$ is much smaller than the electric component and thus 
$\delta \ALP(\x) \approx \delta \ALP_\textrm{eff}(\x)$. Plugging this relation and eq. (\ref{eq:c1c2c3}) into eq.(\ref{eq:ABCD2}), we obtain a quadratic equation in $\beta$ whose positive solution 
\BEA
\hat{\beta}^\textrm{A}& \equiv &\frac{-\hat{b}-\textrm{sgn}\,\hat{c}\sqrt{\hat{b}^2-4 \hat{a} \hat{c}}}{2 \hat{a}}\approx \beta,
\nonumber
\\
\hat{a}&\equiv&2 c_3 \gamma_1+(c_2-c_1)\gamma_2,
\nonumber
\\
\hat{b}&\equiv&(B+C)\gamma_1+(D-A)\gamma_2+(C-B)\kappa,
\nonumber
\\
\hat{c}&\equiv &B-C=-\nabla\times \delta \ALP_{\textrm{eff}},
\label{eq:beta-solution1}
\EEA
is an estimator of $\beta$. If $|\hat{c}|=2\beta|(\gamma_2 \delta \gamma_1-\gamma_1\delta \gamma_2)| \ll \hat{b}^2/|\hat{a}|$ and $\beta >0$, eq. (\ref{eq:beta-solution1}) can be further simplified as 
\BEA
\hat{\beta}^{\textrm{B}} &\equiv& -\frac{\hat{c}}{\hat{b}},
\nonumber
\\
&=& \frac{C-B}{(D-A)\gamma_2+(B+C)\gamma_1+(C-B)\kappa}
\nonumber
\\
&\approx& \beta.
\label{eq:beta-solution2}
\EEA
One can easily confirm that eq. (\ref{eq:beta-solution2}) is equivalent to eq. (\ref{eq:eta})
provided that $\beta c_i$'s are sufficiently small. If not, eq. (\ref{eq:beta-solution1}), which include effects from $\beta c_i$'s may give 
a much better approximation compared to eq. (\ref{eq:beta-solution2}). Note that eqs (\ref{eq:beta-solution1}), (\ref{eq:beta-solution2}) are only valid in some limited regions of $\x$ in which the amplitude of rotation $|\nabla \times \delta \ALP_{\textrm{eff}}|$ is not 'very' small and the strong lensing effect due to the dominant lens or the perturber is not too large. For instance, if $|C-B|$ is sufficiently smaller than the typical 
amplitude of linear perturbations $\delta \kappa, \delta \gamma_1, \delta \gamma_2$, the eqs. (\ref{eq:beta-solution1}), (\ref{eq:beta-solution2})
give a bad approximation. In that case, second order or higher order correction is necessary to give an accurate estimate. If the gradient of a projected gravitational potential of either of the dominant or subdominant lens is too large, eq. (\ref{eq:eta-solution}) gives a bad approximation. In the former case, $c_i$'s become too large. Then the linear approximation in eq. (\ref{eq:c1c2c3}) is no longer valid. In the latter case, the subdominant lens dominates the lensing effect over that of the dominant lens and thus $\delta \alpha$ is no longer smaller than $\alpha$.   

Next, we consider the background perturber case. The components of the perturbed magnification matrix $\delta M_\textrm{eff}$ 
\BEA
A&=& \delta \kappa+\delta \gamma_1-\xi [(\kappa+\gamma_1)(\delta \kappa+\delta \gamma_1)
+\gamma_2 \delta \gamma_2],
\nonumber
\\
B&=& \delta \gamma_2-\xi (\kappa\delta \gamma_2+\gamma_2\delta \kappa
+\gamma_1\delta \gamma_2-\gamma_2\delta \gamma_1),
\nonumber
\\
C&=& \delta \gamma_2-\xi (\kappa\delta \gamma_2+\gamma_2\delta \kappa
+\gamma_2\delta \gamma_1-\gamma_1\delta \gamma_2),
\nonumber
\\
D&=& \delta \kappa-\delta \gamma_1-\xi [(\kappa-\gamma_1)(\delta \kappa-\delta \gamma_1)+\gamma_2 \delta \gamma_2],
\EEA
yield an exact solution
\BE
\xi = \frac{B-C}{(D-A)\gamma_2+(B+C)\gamma_1+(B-C)\kappa}.
\label{eq:xi-solution2}
\EE
If the magnitude of 
rotation $|C-B|=|\hat{c}|$ is sufficiently smaller than $|(D-A)\gamma_2+(B+C)\gamma_1|$, eqs. (\ref{eq:beta-solution2}) and (\ref{eq:xi-solution2}) yield an estimator of $\eta$, 
\BEA
\hat{\eta} &\equiv& \frac{C-B}{(D-A)\gamma_2+(B+C)\gamma_1}
\nonumber
\\
&\approx & \eta.
\label{eq:eta-solution}
\EEA
If $\hat{\eta} $ is positive(negative), it is likely that a perturber resides in the foreground (background).

\subsection{Estimators of LOS perturbations}
In a similar manner, we can express the 'bare'\footnote{Here 'bare' means not influenced by a dominant lens} perturbations $\delta \hat{\kappa}, \delta \hat{\gamma}_1, 
\delta \hat{\gamma}_2$ due to a perturber in the LOS in terms of the derivatives of $\delta \ALP_{\textrm{eff}}$. In the following, we assume that $\beta\ll 1$ and $\xi\ll1$. 

In the foreground perturber case, if $\beta c_i$'s are sufficiently small, we have
\BEA
\delta \hat{\kappa}^{\textrm{B}}&=&F[2(AC+BD)\gamma_1-(A^2+B^2-C^2-D^2)\gamma_2]/2,
\nonumber
\\
\delta \hat{\gamma}_1^{\textrm{B}}&=&F[2(AC-BD)\gamma_1
\nonumber
\\
&-&(A+B-C-D)(A-B+C-D)\gamma_2]/2,
\nonumber
\\
\delta \hat{\gamma}_2^{\textrm{B}}&=&F[2BC\gamma_1-(AB-CD)\gamma_2],
\nonumber
\\
F&\equiv&(\hat{\beta}^{\textrm{B}})^{-1} (C-B)[4 BC \gamma_1^2-2(B+C)(A-D)\gamma_1\gamma_2
\nonumber
\\
&-&(A+B-C-D)(-A+B-C+D)\gamma_2^2 ]^{-1}.
\label{eq:approx-perturbation}
\EEA
If $\beta c_i$'s are not sufficiently small, the approximated 
analytic solutions have a more complicated form (see Appendix). 

Interchanging B and C and substituting $\xi$ into $\beta$ in eq. (\ref{eq:approx-perturbation}),
the exact perturbations for a background perturber are
\BEA
\delta \kappa&=&G[2(AB+CD)\gamma_1-(A^2-B^2+C^2-D^2)\gamma_2]/2,
\nonumber
\\
\delta \gamma_1&=&G[2(AB+CD)\gamma_1
\nonumber
\\
&-&(A+B-C-D)(A-B+C-D)\gamma_2]/2
\nonumber
\\
\delta \gamma_2&=&G[2BC\gamma_1-(AC-BD)\gamma_2],
\nonumber
\\
G&\equiv&\xi^{-1} (B-C)[4 BC \gamma_1^2-2(B+C)(A-D)\gamma_1\gamma_2
\nonumber
\\
&-&(A+B-C-D)(-A+B-C+D)\gamma_2^2 ]^{-1}.
\label{eq:approx-perturbation2}
\EEA

\section{Extended Multi-Plane Mass-sheet Transformation}
In this section, we study how multi-plane mass-sheet transformation (MMST) \cite{schneider2019}
affects the LOS distance ratio and time delay in a double lens system 
with a single source at a certain redshift. A scale transformation in the distance ratio allow a non-zero mass-sheet transformation in both the foreground and background lens planes. This implies that we have two degrees of freedom in scale transformation if the redshift of a perturber is not known. We call such a transformation (multi-plane MST plus a scale transformation in the distance ratio of a perturber) an 'extended MMST' (eMMST). 

\subsection{Distance ratio}
Let us recall mass-sheet transformation (MST) in a single lens system with a single source. 
The position of the source at a certain redshift is given by  
\BE
\y =\x-\ALP(\x).
\label{eq:lens}
\EE
A scale transformation by a factor of $\lambda$ 
\BE
\y \rightarrow \y'=\lambda \y
\label{eq:MST1}
\EE
and a scale transformation by a factor of $\lambda$ 
accompanied by an addition of a deflection by a constant convergence 
\BE
\ALP{(\x)} \rightarrow\ALP'(\x)=\lambda \ALP+(1-\lambda) \bvec{x}
\label{eq:MST2}
\EE
leave a lens equation 
\BE
\y' =\x-\ALP'(\x)
\EE
invariant. The transformed lens system is equivalent to the original lens system(\ref{eq:lens})
except for the physical size and the intensity 
of the source. If we do not know the true size or the 
intensity of the source, we cannot distinguish between the 
two systems with different deflection angles. 
The set of transformations (\ref{eq:MST1}) and  (\ref{eq:MST2})
is called as MST for a single lens system. 
  
In what follows, we consider MMST in a double lens system that 
consists of a dominant lens and a subdominant lens which 
acts as a perturber. 

First, we assume that the subdominant lens with a distance ratio parameter $\beta$ resides in the foreground of the dominant lens. We consider a scale transformation for the dominant lens
\BEA
\y \rightarrow \y'&=&\lambda_\textrm{d} \y
\nonumber
\\
&=& \x-\delta \tilde{\ALP}-(1-\lambda_\textrm{d})\x-\lambda_\textrm{d} \ALP(\x-\beta \delta \ALP)
\label{eq:yprime}
\EEA 
and that for the foreground secondary lens
\BEA
\y_\textrm{d} \rightarrow \y'_\textrm{d}&=&\lambda_\textrm{f} \y_\textrm{d}
\nonumber
\\
&=& \x-(1-\lambda_\textrm{f})\x-\lambda_\textrm{f} \beta \delta \ALP(\x).
\label{eq:ydprime}
\EEA 
If the distance ratio parameter $\beta$ is not known, we need to consider a scale transformation for $\beta$,
\BE
\beta \rightarrow \beta'=\lambda_\beta \beta.
\label{eq:bprime}
\EE
Then eq. (\ref{eq:ydprime}) can be written as
\BEA
\y_\textrm{d} \rightarrow \y'_\textrm{d}&=& \x-\beta' \delta \bvec{\alpha}'(\x)
\nonumber
\\
&=& \x-\beta' \biggl[\frac{1-\lambda_\textrm{f}}{\beta'} \x+ \frac{\lambda_\textrm{f}}{\lambda_\beta} \delta \ALP(\x)  \biggr],
\label{eq:yfprime}
\EEA 
and eq. (\ref{eq:yprime}) as 
\BEA
\y'&=&\x-\delta \tilde{\ALP}-(1-\lambda_\textrm{d})\x-\lambda_\textrm{d} \ALP(\y'_\textrm{d}/\lambda_\textrm{f})
\nonumber
\\
&=& \x-\delta \bvec{\alpha}'(\x)-\tilde{\bvec{\alpha}}(\y'_\textrm{d}(\x,\beta',\delta\bvec{\alpha}'))
\nonumber
\\
&-&\bvec{\varepsilon}(\y'_\textrm{d}(\x,\beta',\delta \bvec{\alpha}')),
\nonumber
\\
\tilde{\bvec{\alpha}}(\y'_\textrm{d}) &\equiv& \lambda_\textrm{d} \ALP(\y'_\textrm{d}/\lambda_\textrm{f}),
\label{eq:yprime2}
\EEA 
where
\BE
\bvec{\varepsilon}(\y'_\textrm{d}(\x,\beta',\delta \bvec{\alpha}'))
\EE
is a residual
function for which $\bvec{\alpha}'\equiv \tilde{\bvec{\alpha}}+\bvec{\varepsilon}$ is a transformed deflection angle. $\bvec{\varepsilon}$ is a function of $\y'_\textrm{d}$ as the transformed lens equation can be expressed as $\y'=\x-\delta \bvec{\alpha}'(\x)-\bvec{\alpha}'(\y'_\textrm{d})$. Here we assume that 
$|\bvec{\varepsilon}|$ is not sufficently smaller than $|\delta \bvec{\alpha}|$. From eq. (\ref{eq:yprime2}), the residual function satisfies
\BE
\bvec{\varepsilon}=\biggl[\frac{-1+\lambda_\textrm{f}}{\beta'}-\lambda_\textrm{d}+1 \biggr]
\x+\biggl[\lambda_\textrm{d}-\frac{\lambda_\textrm{f}}{\lambda_\beta} \biggr] \delta \ALP.
\label{eq:epsilonyd'}
\EE
In terms of $\x$ and $\delta \ALP$, $\bvec{y}'_\textrm{d}$ can be represented as  
\BE
\bvec{y}'_\textrm{d}=\lambda_\textrm{f}(\x-\beta \delta \ALP),
\label{eq:yd'}
\EE
Since $\bvec{\varepsilon}$ is a function of $\bvec{y}'_d$, the ratios of the coefficients of $\x$ and $\delta \ALP$ in eqs. (\ref{eq:epsilonyd'}) and (\ref{eq:yd'}) should be equivalent, and the ratio is equal to $\lambda_\textrm{f}/(\lambda_\textrm{f}\beta)=-\beta^{-1}$. Then we have
\BE
\lambda_\beta=\frac{1}{(1-\lambda_\textrm{d})\beta+\lambda_\textrm{d}}.
\label{eq:lambda_beta}
\EE
If $\beta$ is known, i.e., $\lambda_\beta=1$, then eq. (\ref{eq:lambda_beta}) implies $\lambda_\textrm{d}=1$, which is the known result for MMST in double (main) 
lens systems \cite{schneider2019}. On the contary, if $\beta$ is not known, any value of $\beta'=\lambda_\beta \beta$ can explain most of lensing phenomena by an appropriate MST (specified by $\lambda_\textrm{d}$) for the dominant (background) lens. Thus we have two degrees of freedom among the three scale parameters $\lambda_\textrm{f},\lambda_\textrm{d},$ and $\lambda_\beta$. If $|\bvec{\varepsilon}|$ is sufficiently smaller than 
$|\delta \ALP|$ then $\bvec{\varepsilon}$ is not necessarily a function of
$\bvec{y}'_\textrm{d}$ as $\bvec{\varepsilon}$ is negligible. Therefore, in this case, eq. (\ref{eq:lambda_beta}) no longer holds.
However, from eq. (\ref{eq:epsilonyd'}), which holds even if $\bvec{\varepsilon}$ is not a function of $\y'_\textrm{d}$, one can easily
show that $\lambda_\beta \approx \lambda_\textrm{f} \approx \lambda_\textrm{d} \approx 1$ if $|\bvec{\varepsilon}|\sim 0$. Therefore, we do not have any solutions in which $\beta'$ significantly differs from $\beta$ (i.e., $|1-\lambda_\textrm{d}|$ is not so small with respect to unity) in this case. 
   
Second, we assume that the subdominant lens with a distance ratio parameter $\xi$ resides in the background of the dominant lens. 
We consider a scale transformation for the
dominant lens
\BEA
\y \rightarrow \y'&=&\lambda_\textrm{d} \y
\nonumber
\\
&=& \x-\tilde{\ALP}-(1-\lambda_\textrm{d})\x-\lambda_\textrm{d} \delta \ALP(\x-\xi \delta \ALP),
\label{eq:yprimeb}
\EEA 
that for the background secondary lens
\BEA
\y_\textrm{b} \rightarrow \y'_b&=&\lambda_\textrm{b} \y_\textrm{b}
\nonumber
\\
&=& \x-(1-\lambda_\textrm{b})\x-\lambda_\textrm{b} \xi \delta \ALP(\x),
\label{eq:yprimebg}
\EEA 
and that for the distance ratio parameter
\BE
\xi \rightarrow \xi'=\lambda_\xi \xi.
\label{eq:xiprime}
\EE
By applying a similar argument above to eqs. (\ref{eq:yprimeb}) , (\ref{eq:yprimebg}), and (\ref{eq:xiprime}),  we have
\BE
\lambda_\xi=\frac{1}{(1-\lambda_\textrm{d})\xi+\lambda_\textrm{d}}.
\label{eq:lambda_xi}
\EE
Thus the mass-sheet degeneracy (MSD) in the distance ratios $\beta$ and $\xi$, which is related to the scale transformation in the background lens remain.

\subsection{Time delay}
%Assuming that the Hubble constant is known, measurement o
%f time delays between multiple lensed images of a compact source and their astrometric lensing B-modes% could break the degeneracy in the distance ratios of dark haloes that affect the lensed images. 
\begin{figure}
\center
\includegraphics[width=8cm,pagebox=cropbox,clip]{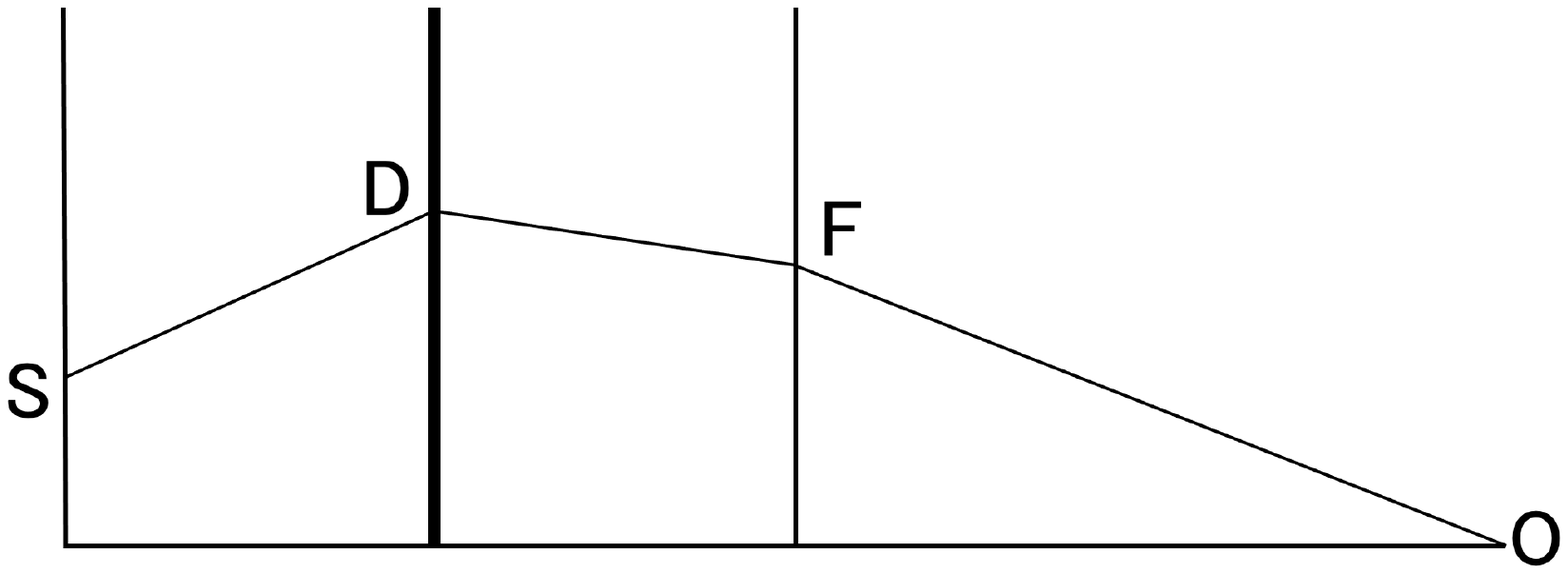}
\vspace{-0.8cm}
\caption{An example of a light path for a foreground perturber. A light ray emitted from a point source at S 
in the source plane is deflected by a dominant lens at D and a subdominant lens at F and reaches to an obserber at O.   }
\label{fig:background-tau}
\end{figure} 
First we examine a double lens system with a foreground perturber. 
Then the time delay $\tau $ of a source at $\bvec{y}$, which is defined as 
the arrival time difference between a light path that is lensed by a dominant lens at known redshift $z_\textrm{d}$ and a foreground perturber at unknown redshift $z_\textrm{f}$ and an unlensed path is
\BEA
c \tau&=& c \tau_{\textrm{f}}+c\tau_{\textrm{d}}
\nonumber
\\
&=&
(1+z_\textrm{f}) \frac{D_{\textrm{f}}D_{\textrm{d}}}{D_{\textrm{df}} }
\biggl[-\nabla^{-1}_\x \bigl(\beta \delta \ALP(\x) \bigr)+\frac{(\x-\y_\textrm{d})^2}{2}  \biggr]
\nonumber
\\
&+&(1+z_\textrm{d}) \frac{D_{\textrm{d}}D_{\textrm{s}}}{D_{\textrm{sd}} }
\biggl[-\nabla^{-1}_{\y_\textrm{d}} \bigl(\ALP(\y_\textrm{d}) \bigr)+\frac{(\y_\textrm{d}-\y)^2}{2}  \biggr],
\nonumber
\\
\label{eq:tau}
\EEA
where $\tau_{\textrm{f}}$ and $\tau_{\textrm{d}}$ are the arrival time differences between light paths DFO and DO, and SDO and SO, respectively (as depicted in Fig. \ref{fig:background-tau}). The constant $c$ is the light speed. We investigate how the time delay $\tau$ behaves under eMMST. The scalings (\ref{eq:yprime}), (\ref{eq:ydprime}), and (\ref{eq:bprime}) yield
\BEA
c \tau'&=& c \tau'_{\textrm{f}}+c\tau'_{\textrm{d}}
\nonumber
\\
&=&
(1+z'_\textrm{f}) \frac{D'_{\textrm{f}}D_{\textrm{d}}}{D'_{\textrm{df}} }
\biggl[-\nabla^{-1}_\x \bigl(\beta' \delta \ALP'(\x) \bigr)+\frac{(\x-\y'_\textrm{d})^2}{2}  \biggr]
\nonumber
\\
&+&(1+z_\textrm{d}) \frac{D_{\textrm{d}}D_{\textrm{s}}}{D_{\textrm{sd}} }
\biggl[-\nabla^{-1}_{\y'_\textrm{d}} \bigl(\ALP'(\y'_\textrm{d}) \bigr)+\frac{(\y'_\textrm{d}-\y')^2}{2}  \biggr]
\nonumber
\\
&=& 
(1+z'_\textrm{f}) \frac{D'_{\textrm{f}}D_{\textrm{d}}}{D'_{\textrm{df}} }
\biggl[-\lambda_\textrm{f} \nabla^{-1}_\x \bigl(\beta \delta \ALP(\x) \bigr)-
\frac{(1-\lambda_\textrm{f})x^2}{2}
\nonumber
\\
&+&\frac{(\x-\lambda_\textrm{f} \y_\textrm{d})^2}{2}  \biggr]+ (1+z_\textrm{d}) \frac{D_{\textrm{d}}D_{\textrm{s}}}{D_{\textrm{sd}} }\biggl[-\lambda_\textrm{f} \lambda_\textrm{d} \nabla^{-1}_{\y_\textrm{d}} \bigl(\ALP(\y_\textrm{d}) \bigr)
\nonumber
\\
&-&\frac{\lambda_\textrm{f}^2 h \,y^2_\textrm{d}}{2}+\frac{(\lambda_\textrm{d}\y_\textrm{d}-\lambda_\textrm{f}\y)^2}{2}  \biggr],
\nonumber
\\
h &\equiv& - \frac{\lambda_\textrm{d}}{\beta \lambda_\textrm{f}}+\frac{1}{\lambda_\beta \beta}.
\label{eq:taup}
\EEA
Since
\BEA
& &(1+z'_\textrm{f}) \frac{D'_{\textrm{f}}D_{\textrm{d}}}{D'_{\textrm{df}} }
\biggl[- \frac{(1-\lambda_\textrm{f})x^2}{2}+\frac{(\x-\lambda_\textrm{f} \y_\textrm{d})^2}{2}  \biggr]
\nonumber
\\
&+&(1+z_\textrm{d}) \frac{D_{\textrm{d}}D_{\textrm{s}}}{D_{\textrm{sd}} }\biggl[-\frac{\lambda_\textrm{f}^2 h \,y^2_\textrm{d}}{2}+\frac{(\lambda_\textrm{d}\y_\textrm{d}-\lambda_\textrm{f}\y)^2}{2} \biggr]
\nonumber
\\
&-&(1+z'_\textrm{f}) \frac{D'_{\textrm{f}}D_{\textrm{d}}}{D'_{\textrm{df}} }
\frac{\lambda_\textrm{f}(\x-\y_\textrm{d})^2}{2}  
\nonumber
\\
&-&(1+z_\textrm{d}) \frac{D_{\textrm{d}}D_{\textrm{s}}}{D_{\textrm{sd}} }
\frac{\lambda_\textrm{d} \lambda_\textrm{f} (\y_\textrm{d}-\y)^2}{2} 
\nonumber
\\
&=& (1+z'_\textrm{f}) \frac{D'_{\textrm{f}}D_{\textrm{d}}}{D'_{\textrm{df}} }
\frac{\lambda_\textrm{d}(\lambda_\textrm{d}-\lambda_\textrm{f})y^2}{2}
\nonumber
\\
&+&(1+z_\textrm{d}) \frac{D_{\textrm{d}}D_{\textrm{s}}}{D_{\textrm{sd}} }
\frac{\lambda_\textrm{f}(\lambda_\textrm{f}-1)(\lambda_\textrm{d}-1)y^2_\textrm{d}}{2},
\EEA
and
\BE
(1+z'_\textrm{f}) \frac{D'_{\textrm{f}}D_{\textrm{d}}}{D'_{\textrm{df}}}=-(1+z_\textrm{d}) \frac{D_{\textrm{d}}D_{\textrm{s}}}{D_{\textrm{sd}} }
\lambda_\textrm{d} \biggl[1-\frac{1}{\beta}\biggr],
\EE
eq. (\ref{eq:taup}) gives
\BEA
c \tau'&=& \lambda_\textrm{f} \lambda_\textrm{d} c \tau
+ \frac{\lambda_\textrm{d}(\lambda_\textrm{d}-\lambda_\textrm{f})F_\textrm{sd} y^2}{2},
\nonumber
\\
F_\textrm{sd}&\equiv& (1+z_\textrm{d}) \frac{D_{\textrm{d}}D_{\textrm{s}}}{D_{\textrm{sd}} }.
\label{eq:taup2}
\EEA
Since the last term in eq. (\ref{eq:taup2}) depend only on the position of a source, it does not contribute to time delay between lensed images. Thus eMMST admits a scale transformation in the foreground lens plane and another one in the background lens plane. Let us suppose that quadruply lensed images consist of two images that are lensed by only a dominant lens and the other two images that are perturbed by a foreground halo. Then the time delay between unperturbed images gives $\lambda_\textrm{d}$ and that between perturbed images gives $\lambda_\textrm{f}$ given a Hubble constant $H_0$. 

\begin{figure}
\center
\includegraphics[width=8cm,pagebox=cropbox,clip]{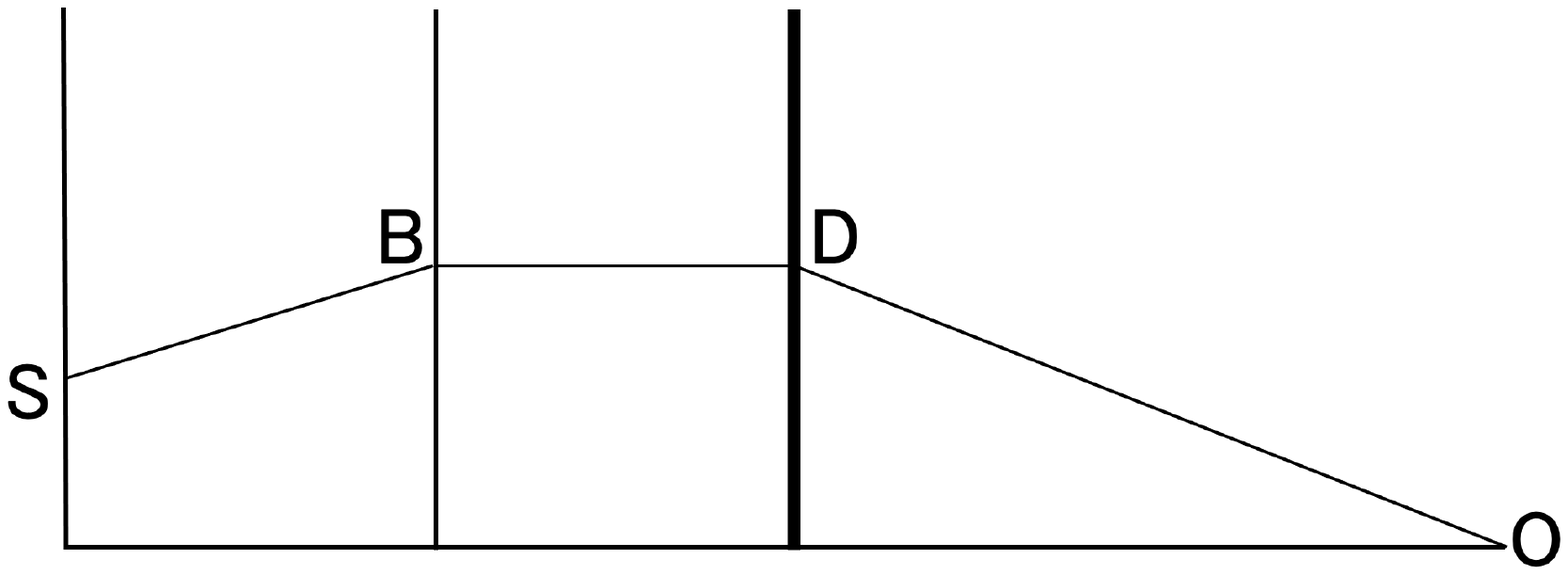}
\vspace{-0.8cm}
\caption{An example of a light path for a background perturber. A light ray emitted from a point source at S 
in the source plane is deflected by a subdominant lens at B and a dominant lens at F and reaches to an obserber at O.   }
\label{fig:foreground-tau}
\end{figure} 

Next we examine a double lens system with a background perturber. In this case, the time delay of a source at $\y$ is
\BEA
c \tau&=& c \tau_{\textrm{d}}+c\tau_{\textrm{b}}
\nonumber
\\
&=&
(1+z_\textrm{d}) \frac{D_{\textrm{d}}D_{\textrm{b}}}{D_{\textrm{bd}} }
\biggl[-\nabla^{-1}_\x \bigl(\xi \ALP(\x) \bigr)+\frac{(\x-\y_\textrm{b})^2}{2}  \biggr]
\nonumber
\\
&+&(1+z_\textrm{b}) \frac{D_{\textrm{b}}D_{\textrm{s}}}{D_{\textrm{sb}} }
\biggl[-\nabla^{-1}_{\y_\textrm{b}} \bigl(\delta \ALP(\y_\textrm{d}) \bigr)+\frac{(\y_\textrm{b}-\y)^2}{2}  \biggr],
\nonumber
\\
\label{eq:taub}
\EEA
where $\tau_{\textrm{d}}$ and $\tau_{\textrm{b}}$ are the arrival time differences between light paths BDO and BO, and SBO and SO, respectively (as depicted in Fig. \ref{fig:foreground-tau}). By applying a similar argument above, the scalings (\ref{eq:yprimeb}), (\ref{eq:yprimebg}), and (\ref{eq:xiprime}) yield 
\BEA
c \tau'&=& c \tau'_{\textrm{d}}+c\tau'_{\textrm{b}}
\nonumber
\\
&=&
(1+z_\textrm{d}) \frac{D_{\textrm{d}}D'_{\textrm{b}}}{D'_{\textrm{bd}} }
\biggl[-\nabla^{-1}_\x \bigl(\xi' \ALP'(\x) \bigr)+\frac{(\x-\y'_\textrm{b})^2}{2}  \biggr]
\nonumber
\\
&+&(1+z'_\textrm{b}) \frac{D'_{\textrm{b}}D_{\textrm{s}}}{D'_{\textrm{sb}} }
\biggl[-\nabla^{-1}_{\y'_b} \bigl(\delta \ALP'(\y'_b) \bigr)+\frac{(\y'_b-\y')^2}{2}  \biggr]
\nonumber
\\
&=& 
(1+z_\textrm{d}) \frac{D_{\textrm{d}}D'_{\textrm{b}}}{D'_{\textrm{bd}} }
\biggl[-\lambda_\textrm{d} \nabla^{-1}_\x \bigl(\xi \ALP(\x) \bigr)-
\frac{(1-\lambda_\textrm{d})x^2}{2}
\nonumber
\\
&+&\frac{(\x-\lambda_\textrm{d} \y_\textrm{b})^2}{2}  \biggr]+ (1+z'_\textrm{b}) \frac{D'_{\textrm{b}}D_{\textrm{s}}}{D'_{\textrm{sb}} }\biggl[-\lambda_\textrm{d} \lambda_\textrm{b} \nabla^{-1}_{\y_\textrm{b}} \bigl(\delta \ALP(\y_\textrm{b}) \bigr)
\nonumber
\\
&-&\frac{\lambda_\textrm{d}^2 g \,y^2_b}{2}+\frac{(\lambda_\textrm{d}\y_\textrm{b}-\lambda_\textrm{b}\y)^2}{2}  \biggr],
\nonumber
\\
g &\equiv& - \frac{\lambda_\textrm{b}}{\xi \lambda_\textrm{d}}+\frac{1}{\lambda_\xi \xi}
\label{eq:taupb}
\EEA
Since
\BEA
& &(1+z_\textrm{d}) \frac{D_{\textrm{d}}D'_{\textrm{b}}}{D'_{\textrm{bd}} }
\biggl[- \frac{(1-\lambda_\textrm{d})x^2}{2}+\frac{(\x-\lambda_\textrm{d} \y_\textrm{b})^2}{2}  \biggr]
\nonumber
\\
&+&(1+z'_\textrm{b}) \frac{D'_{\textrm{b}}D_{\textrm{s}}}{D'_{\textrm{sd}} }\biggl[-\frac{\lambda_\textrm{d}^2 g \,y^2_b}{2}+\frac{(\lambda_\textrm{d}\y_\textrm{b}-\lambda_\textrm{b}\y)^2}{2} \biggr]
\nonumber
\\
&-&(1+z_\textrm{d}) \frac{D_{\textrm{d}}D'_{\textrm{b}}}{D'_{\textrm{bd}} }
\frac{\lambda_\textrm{d}(\x-\y_\textrm{b})^2}{2}  
\nonumber
\\
&-&(1+z'_\textrm{b}) \frac{D'_{\textrm{b}}D_{\textrm{s}}}{D'_{\textrm{sd}} }
\frac{\lambda_\textrm{b} \lambda_\textrm{d} (\y_\textrm{b}-\y)^2}{2} 
\nonumber
\\
&=& (1+z'_\textrm{b}) \frac{D'_{\textrm{b}}D_{\textrm{s}}}{D'_{\textrm{sb}} }
\frac{\lambda_\textrm{b}(\lambda_\textrm{b}-\lambda_\textrm{d})y^2}{2}
\nonumber
\\
&+&(1+z'_\textrm{b}) \frac{D'_{\textrm{b}}D_{\textrm{s}}}{D'_{\textrm{sb}} }
\frac{\lambda_\textrm{b}\lambda_\textrm{d} (\lambda_\textrm{d}-1)(1-\xi^{-1})y_\textrm{b}^2}{2}
\nonumber
\\
&+&(1+z_\textrm{d}) \frac{D_{\textrm{d}}D'_{\textrm{b}}}{D'_{\textrm{bd}} }
\frac{\lambda_\textrm{d} (\lambda_\textrm{d}-1)y^2_b}{2}
\EEA
and
\BEA
(1+z_\textrm{d}) \frac{D_{\textrm{d}}D'_{\textrm{b}}}{D'_{\textrm{bd}}}&=&(1+z_d) \frac{D_s D_d}{D_{sd}} \frac{1}{\lambda_\xi \xi},
\nonumber
\\
(1+z'_\textrm{b}) \frac{D'_{\textrm{b}}D_{\textrm{s}}}{D'_{\textrm{sb}}}&=&-(1+z_d) \frac{D_s D_d}{D_{sd}} \frac{1}{\lambda_\xi\xi-1},
\EEA
eq. (\ref{eq:taupb}) gives
\BEA
c \tau'&=& \frac{\lambda_\textrm{d}}{\lambda_\xi} c \tau
+ \frac{\lambda_\textrm{b}(\lambda_\textrm{b}-\lambda_\textrm{d})F'_\textrm{sb} y^2}{2},
\nonumber
\\
F'_\textrm{sb}&\equiv& (1+z'_\textrm{b}) \frac{D'_{\textrm{b}}D_{\textrm{s}}}{D'_{\textrm{sb}} },
\label{eq:taupb2}
\EEA
where $\lambda_\xi$ is given by eq. (\ref{eq:lambda_xi}).
Since the last term in eq. (\ref{eq:taupb2}) depends solely on the position of the source, it does not contribute to the time delay between lensed images. In the case of background perturbers, eMMST admits a scale transformation in the foreground lens plane and another one in the background lens plane. However, the transformed time delay between lensed images is inversely proportional to the scale factor $\lambda_\xi$. If $\lambda_\textrm{d}$ is determined by the observation of time delays between unperturbed images, it is possible to measure $\lambda_\xi$ with astrometric lensing B-mode for an assumed Hubble constant $H_0$.

In scenarios where the redshifts of both foreground and background perturbers, as well as the dominant lens and the source, are known, observations of astrometric lensing B-mode can break the MSD. This leads to a reduction in systematic errors in our estimated value of $H_0$. The reason is as follows: Let's consider a scenario where a quadruple lens system, with images A, B, C, and D, is perturbed by a foreground perturber affecting image A and a background perturber affecting image B. We assume that their gravitational effects are confined to the vicinity of lensed images A and B, respectively. Additionally, we assume that the gravitational influence of all other perturbers, apart from the dominant lens, can be considered negligible.In this context, a measurement of astrometric lensing B-mode in the de-lensed image A can provide us with $\lambda_\textrm{d}$, as we already have knowledge of $\beta$. Similarly, a measurement of astrometric lensing B-mode in the de-lensed image B can furnish us with either $\lambda_\xi$ or $\lambda_\textrm{b}$ since $\xi$ is a known quantity. Subsequently, by measuring the time delay between BC or BD, we can determine $H_0$, while measuring the time delay between AC or AD allows us to ascertain $\lambda_\textrm{f}$.

\section{Toy models}
We investigate the property of astrometric lensing B-mode and the accuracy of 
the approximated distance ratios $\hat{\beta}^\textrm{A}, \hat{\beta}^\textrm{B},$ and $\hat{\eta}$ for a foreground perturber using simple toy models. As a model of a galaxy halo, we adopt an singular isothermal sphere (SIS) whose reduced deflection angle is given by $\ALP(\x)=\theta_\textrm{E} \x/x$.

One can easily show that any constant shift perturbation in a secondary lens plane can be explained by a translation in the source plane without a secondary lens. Since we do not have any information about the original position of the source, we cannot measure a constant shift. Moreover, a constant external convergence perturbation cannot be measured due to the eMMST. Therefore, in order to probe the effects of large-scale perturbation, we analyse the dominant SIS with a constant external shear at an arbitrary redshift. In order to probe the effects of small-scale perturbation, we analyse the dominant SIS with another subdominant SIS with an Einstein radius of $\theta_\textrm{{Ep}}\ll \theta_\textrm{E}$ that acts as a perturber at an arbitrary redshift. 
\begin{figure*}
\center
\vspace{-3cm}
\includegraphics[width=15.5cm,pagebox=cropbox,clip]{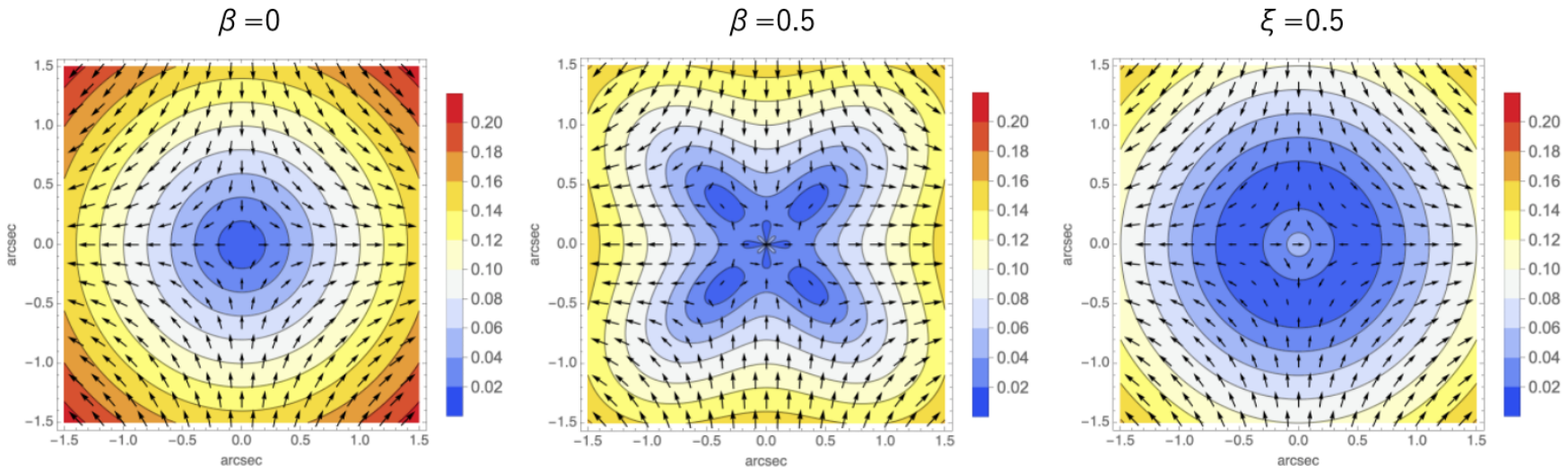}
\vspace{-3.2cm}
\caption{Line-of-sight effect for an SIS model with an external constant shear in the shear-aligned coordinates. The dominant lens is an SIS with Einstein radius $\theta_\textrm{E}=1\,$arcsec centered at $(0,0)$. The arrows show the effective deflection angle $\delta \ALP_{\textrm{eff}}$ due to an external shear with $\delta \gamma=0.1$. The colors show the amplitudes of the shift in arcsec.  }
\label{fig:SISES-shift}
\end{figure*} 

\begin{figure}
\center
\hspace{-0.8cm}
\includegraphics[width=13.cm,pagebox=cropbox,clip]{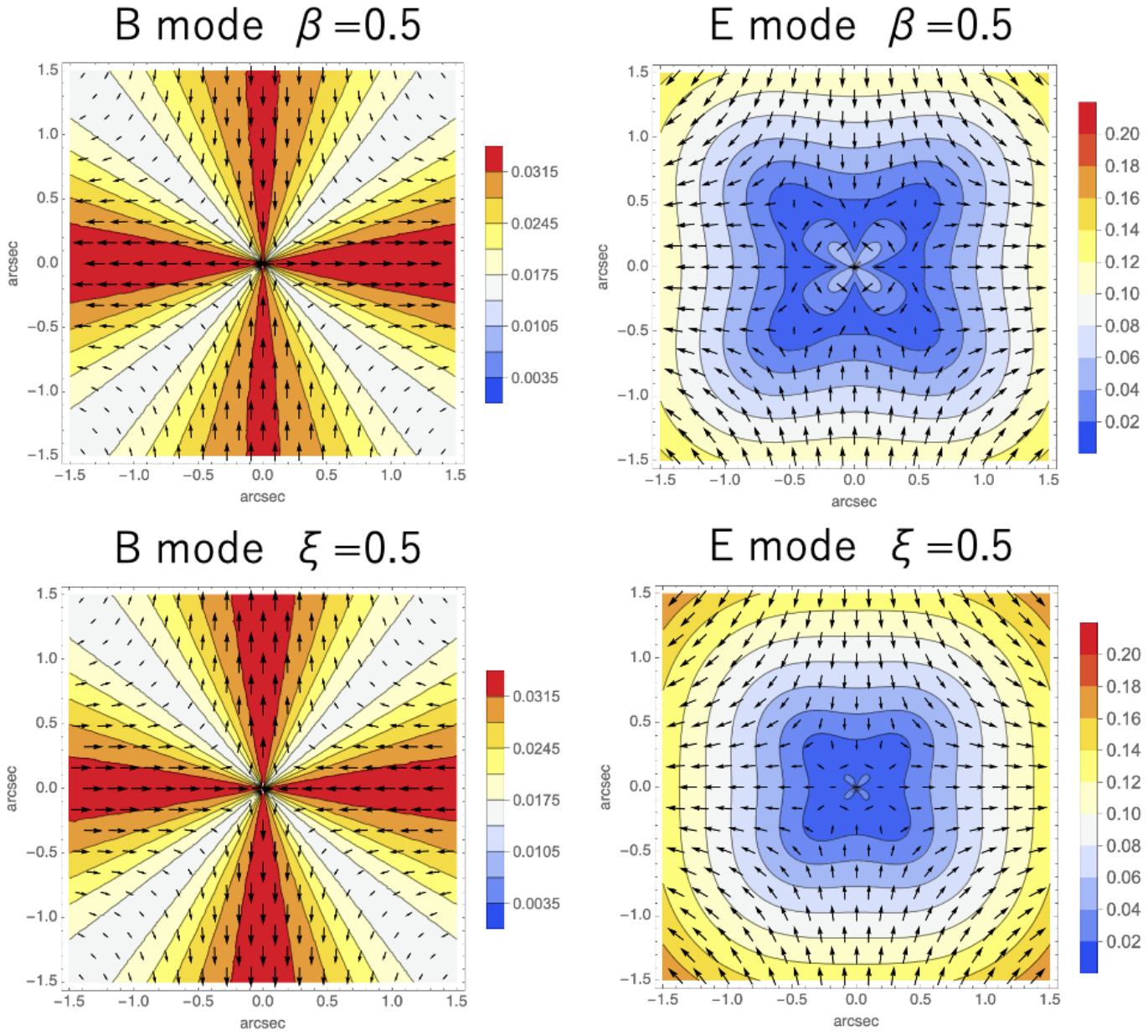}
\vspace{0.5cm}
\caption{Astrometric lensing B and E-modes for an external shear of $\delta \gamma=0.1$. The arrows show the B and E components in the effective astrometric shift $\delta \ALP_{\textrm{eff}}$. The model parameters are the same as in Fig. \ref{fig:SISES-shift}. The colors show the amplitudes of the shift in arcsec.           }
\label{fig:SISES-shift-EB}
\end{figure} 
\subsection{SIS + external shear}
Here we study a lens model with a dominant SIS and a constant external shear $\delta \gamma$. We assume that the center of the effective gravitational potential of the dominant SIS and that of the constant shear coincide. If a perturber resides in the foreground of a dominant SIS, the effective deflection angle of the perturber is 
\BE
\delta \ALP_{\textrm{eff}}^\textrm{f}(\x)=\delta \gamma \mathcal{I}\x +\theta_\textrm{E}\frac{\x-\beta \delta \gamma \mathcal{I}\x}{\lVert\x-\beta \delta \gamma \mathcal{I}\x\rVert}-\theta_\textrm{E} \frac{\x}{x},
\EE
where $\mathcal{I}$ is a unit matrix. If a perturber resides in the background of a dominant SIS, the effective deflection angle of the perturber is
\BE
\delta \ALP_{\textrm{eff}}^\textrm{b}(\x)=\delta \gamma \mathcal{I} \biggl[\x-\xi \theta_\textrm{E} \frac{\x}{x}\biggr].
\EE
The 'magnetic' potential $\psi^\textrm{B}$ can be obtained by solving the Poisson equation 
\BE
-\Delta \psi^\textrm{B}=\nabla \times \delta \ALP_{\textrm{eff}}^\textrm{f}(\x).
\label{eq:poisson}
\EE
To solve eq. (\ref{eq:poisson}) numerically, we used the 
Finite Element Method (FEM). We discretized the inner region of a disk with a radius of $\theta=100\,\theta_\textrm{E}$ into $\sim 2\times 10^5$ triangle meshes in the polar coordinates $(\theta, \phi)$. We used finer meshes in regions with $\theta<2\,\theta_\textrm{E}$ in order to resolve sudden changes in the perturbed magnetic potential. We imposed a Dirichlet boundary condition at $\theta=100\,\theta_\textrm{E}$. The amplitude of rotation can be estimated as $|\nabla \times \delta \ALP_{\textrm{eff}}^\textrm{f} | \sim \beta \delta \gamma/\theta $ or $|\nabla \times \delta \ALP_{\textrm{eff}}^\textrm{b} | \sim \xi \delta \gamma/\theta $. Then, the amplitude at $\theta=100\,\theta_\textrm{E}$ is expected to be $\lesssim 10^{-3}$ for $\delta \gamma \le 0.1$. We numerically checked that the tiny error at the boundary of the disk does not significantly affect the accuracy of the obtained solution at $\theta \lesssim \theta_\textrm{E}$. In order to avoid a singularity at the center of the dominant lens, we used a cored isothermal sphere with very small core radius $\theta_\textrm{c}\ll \theta_\textrm{E}$. The effective deflection angle is given by $\ALP(\x)=\theta_\textrm{E} (\sqrt{x^2+\theta_\textrm{c}^2}-\theta_\textrm{c}) \x/x$. Except for the neighbourhood of singular points (centres of SIS), the absolute errors in the Poisson equation (\ref{eq:poisson}) for numerically obtained solutions of astrometric lensing B-mode $\delta \ALP^{\textrm{B}}_{\textrm{eff}}$ were typically $\lesssim 10^{-2}$.          

The effective deflection angles $\delta \ALP_{\textrm{eff}}$ in an SIS with an external shear of $\gamma\!=\!0.1$ are shown in Fig. \ref{fig:SISES-shift}. If objects that causes an external shear reside in the foreground(background) of a dominant lens, the constant shear produces a cross shape (ring shape) pattern in the field of $\lVert \delta \ALP_{\textrm{eff}} \rVert$. Except for the central region in the center of an SIS, the constant shear reduces the amplitude of $\delta \ALP_{\textrm{eff}}$. The amplitudes of B-mode $|\delta \ALP_{\textrm{eff}}^{\textrm{B}}|$ is as large as $\sim 0.03$\,arcsec in the vicinity of coordinate axes and negligibly small in the vicinity of diagonal lines if $\beta=0.5$ or $\xi=0.5$. The direction of $\delta \ALP_{\textrm{eff}}^{\textrm{B}}$ in the foreground case is opposite to that in the background case. In contrast, the amplitudes of E-mode $|\delta \ALP_{\textrm{eff}}^{\textrm{E}}|$ are largest in the diagonal lines and smallest in the coordinate axes except for the central region (Fig. \ref{fig:SISES-shift-EB}). For a given distance from the center, the amplitudes of rotation $\nabla \times \delta \ALP_{\textrm{eff}}^{\textrm{B}}$ are largest in the diagonal lines (Fig. {fig:SISES-rot}). The 'direction' of rotation in the foreground case is opposite to that in the background case. From the 'direction' of rotation, we can determine whether the perturber resides in the foreground or background of the dominant lens.          

\begin{figure}
\center
\vspace{-1cm}
\includegraphics[width=10cm,pagebox=cropbox,clip]{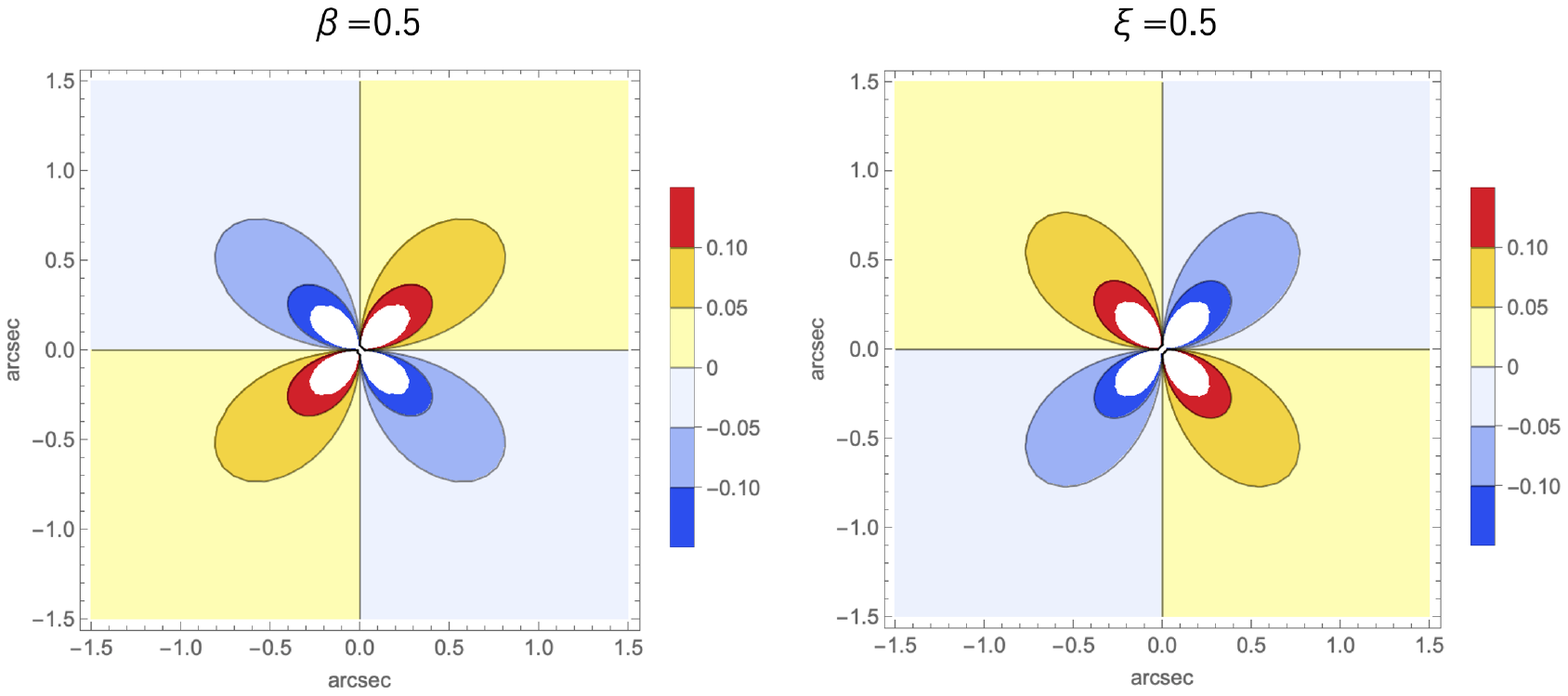}
\vspace{-1cm}
\caption{Rotation of effective deflection angle for an external shear of $\delta \gamma=0.1$. The colors show the amplitudes of rotation. The model parameters are the same as in Fig. \ref{fig:SISES-shift}.  }
\label{fig:SISES-rot}
\end{figure} 
\begin{figure*}
\center
\vspace{-3cm}
\includegraphics[width=15.5cm,pagebox=cropbox,clip]{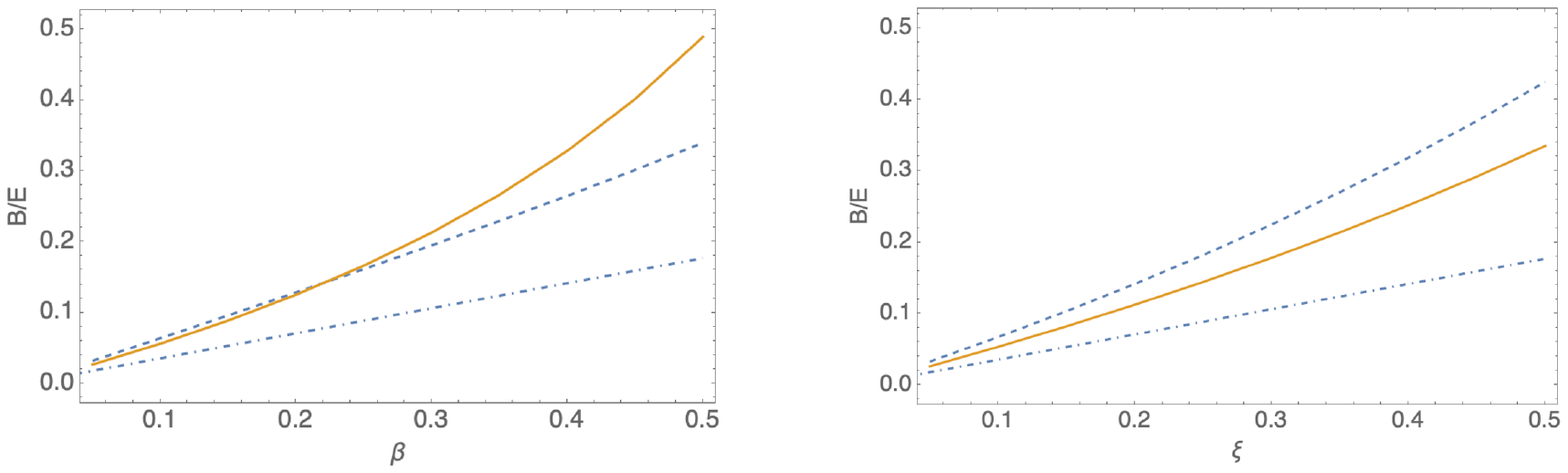}
\vspace{-3cm}
\caption{The ratio of B and E-modes for an external shear of $\delta \gamma=0.1$ and $\theta_\textrm{E}=1\,$arcsec. The orange full curves denote the ratio of the averaged amplitude of B-mode astrometric shift and that of E-mode astrometric shift as a function of $\beta$ (top) and $\xi$ (bottom). Both the modes are averaged over a circle with $\theta=\theta_\textrm{E}$. The blue dashed curves show the ratio of averaged rotation and averaged convergence. Both the two contributions are averaged over a circle with $\theta=\theta_\textrm{E}$. The blue dot-dashed curves represent the approximations of the ratios for the foreground case $\sqrt{2}\beta/4$ (top) and the background case $\sqrt{2}\xi/4$ (bottom).    }
\label{fig:SISES-beta-approx-ave}
\end{figure*} 
\begin{figure*}
\center
\vspace{-3cm}
\hspace{0.cm}
\includegraphics[width=15.5cm,pagebox=cropbox,clip]{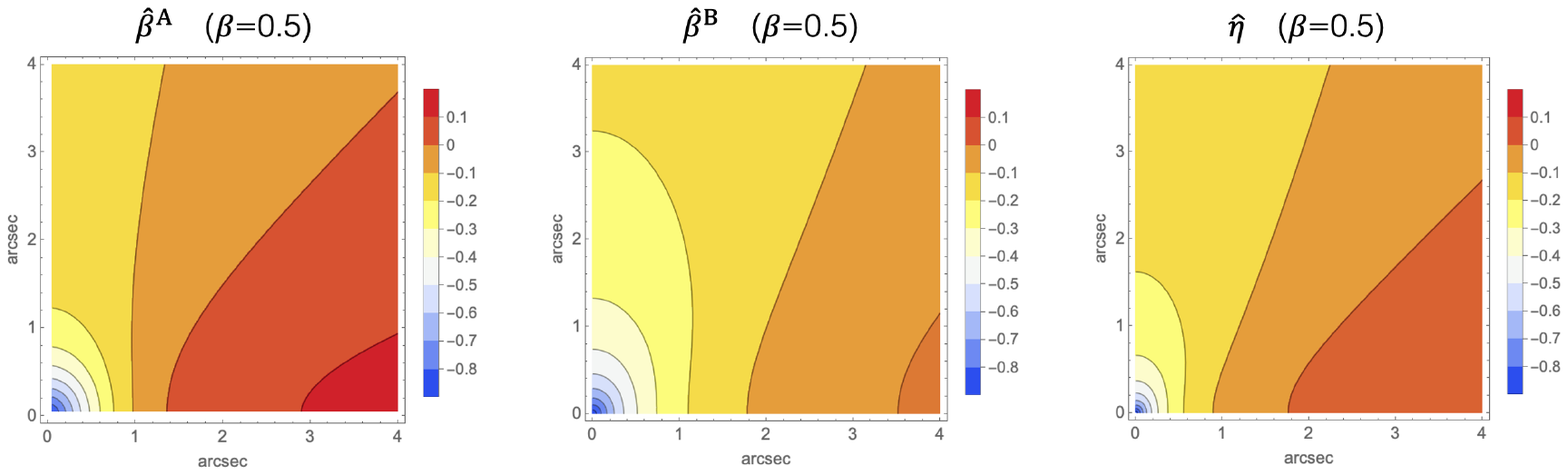}
\vspace{-3.5cm}
\caption{Relative errors of approximated distance ratios in the lens plane for $\delta \gamma=0.1$, $\beta=0.5$, and $\theta_\textrm{E}=1\,$arcsec. The colors show the relative errors 
$\hat{\beta}^{\textrm{A}}/\beta-1$ (left), $\hat{\beta}^{\textrm{B}}/\beta-1$ (middle), and $\hat{\eta}/\beta-1$ (right). }
\label{fig:SISES-beta-approx}
\end{figure*} 

\begin{figure*}
\center
\vspace{-3cm}
\hspace{-0.5cm}
\includegraphics[width=15.5cm,pagebox=cropbox,clip]{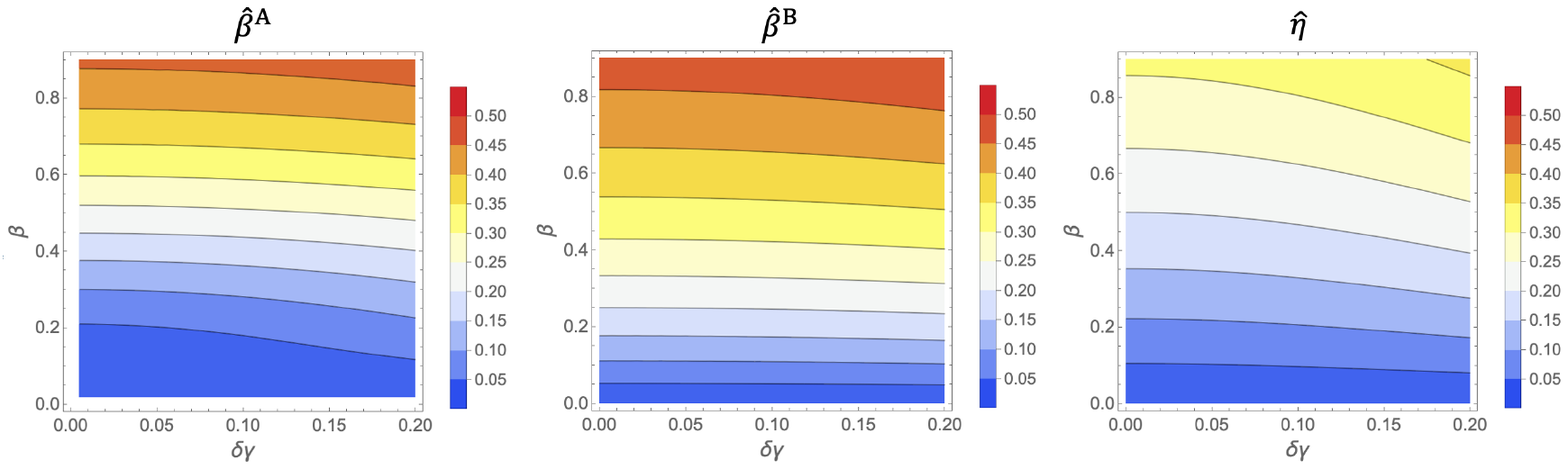}
\vspace{-3cm}
\caption{Relative errors of approximated distance ratio as a function of $\beta$ and $\delta \gamma$. The colors show the relative errors averaged over an Einstein ring with radius $\theta_\textrm{E}$ for a given set of $\delta \gamma$ and $\beta$. }
\label{fig:SISES-beta-approx-deltagamma}
\end{figure*} 

In Fig. \ref{fig:SISES-beta-approx-ave}, we show the ratio of the B-mode to E-mode $r^{\textrm {BE}}=|\delta \ALP_{\textrm{eff}}^{\textrm{B}} |/|\delta \ALP_{\textrm{eff}}^{\textrm{E}}|$, averaged over the Einstein ring with radius $\theta_\textrm{E}=1\,$arcsec. The ratio $r^{\textrm{BE}}$ is a monotonically increasing function of $\beta$ or $\xi$, which equals $\sim 0.5$ for $\beta=0.5$, and $\sim 0.3$ for $\xi=0.5$. We found that the ratio of rotation to divergence, $|\nabla \times \delta \ALP_{\textrm{eff}}^{\textrm{B}} |/|\nabla \cdot \delta \ALP_{\textrm{eff}}^{\textrm{E}}|$ averaged over the Einstein ring gives a good approximation of $r^{\textrm{BE}}$. The analytic formula in eq. (\ref{eq:etaest2}) gives a good estimate for small distance ratio $\beta \lesssim 0.1$ or $\xi \lesssim 0.1$, but the difference is conspicuous for $\beta \gtrsim 0.1$ or $\xi \gtrsim 0.1$. 
 Next, we show the relative errors of the distance ratio estimators $\hat{\beta}^{\textrm{A}}$, $\hat{\beta}^{\textrm{B}}$, and $\hat{\eta}$ in the dominant lens plane in Fig. \ref{fig:SISES-beta-approx}. We found that these estimators give a worse approximation in the central region within the Einstein radius $\theta_\textrm{E}$ and a good approximation in the outer regions $\theta>\theta_\textrm{E}$, especially in the vicinity of the horizontal axis $x_1$. At $\theta=\theta_\textrm{E}=1\,$arcsec, for $\beta \lesssim 0.4$, $\hat{\beta}^{\textrm{A}}$ gives the best approximation but for $\beta \gtrsim 0.4$, $\hat{\eta}$ gives the best approximation (Fig. \ref{fig:SISES-beta-approx-deltagamma}). Thus, a hybrid use of $\hat{\beta}^{\textrm{A}}$ and $\hat{\eta}$ may be a best way to estimate the distance ratio $\beta$ in the SIS + external shear model. 

\subsection{SIS + SIS}
Here we study lens models with a dominant SIS with an Einstein radius of $\theta_\textrm{E}=1\,$arcsec and a subdominant SIS with an Einstein radius of $\theta_\textrm{E0}\ll \theta_\textrm{E}$ apparently centred at $\x_0=(x_1,x_2)=(1'',0)$. If a subdominant SIS resides in the foreground of the dominant SIS, the effective deflection angle of the perturber is 
\BEA
\delta \ALP_{\textrm{eff}}^\textrm{f}(\x)
&=&\theta_\textrm{E0}\frac{\x-\x_{0}}{\lVert \x-\x_{0} \rVert}
+\theta_\textrm{E}\biggl(\frac{\y_\textrm{d}}{y_\textrm{d}}- \frac{\x}{x}\biggr),
\nonumber
\\
\y_\textrm{d}&=&\x-\beta \theta_\textrm{E0} \frac{\x-\x_{0}}{\lVert \x-\x_{0} \rVert}.
\EEA
If an SIS resides in the background of a dominant SIS, the effective deflection angle of the perturber is
\BEA
\delta \ALP_{\textrm{eff}}^\textrm{b}(\x)
&=& \theta_\textrm{E0}\frac{\y_\textrm{b}-\y_\textrm{b0}}{\lVert \y_\textrm{b}-\y_\textrm{b0}\rVert },
\nonumber
\\
\y_\textrm{b}&=&\x-\xi \theta_\textrm{E}\frac{\x}{x},
\nonumber
\\
\y_\textrm{b0}&=&\x_0-\xi \theta_\textrm{E}\frac{\x_0}{x_0}.
\EEA

\begin{figure*} \center
\vspace{-3.5cm}
\includegraphics[width=15.5cm,pagebox=cropbox,clip]{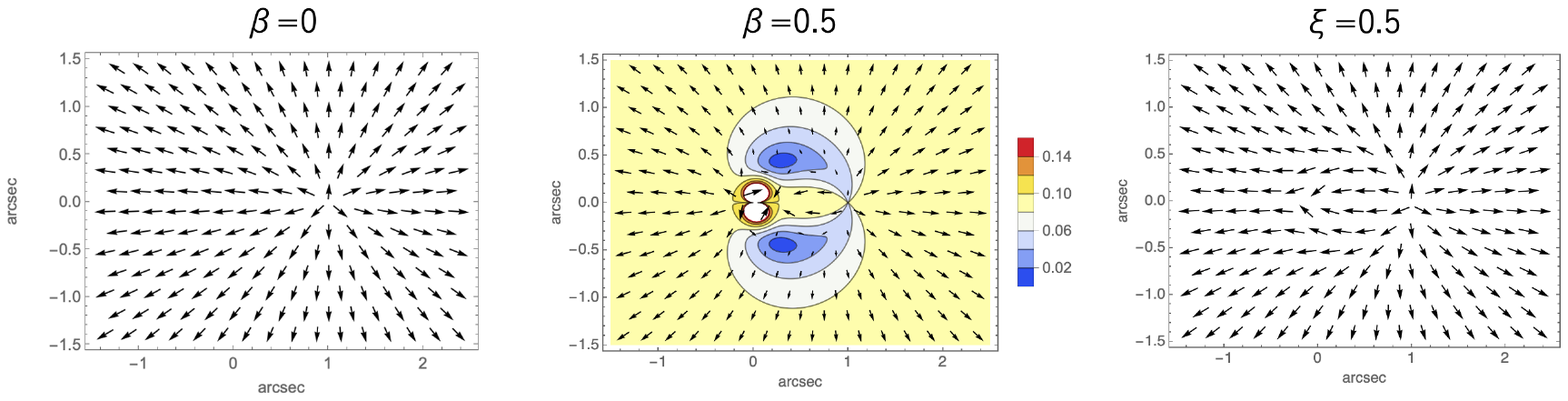}
\vspace{-4cm}
\caption{Line-of-sight effect for an SIS+SIS model. The dominant lens is an SIS with Einstein radius $\theta_\textrm{E}=1\,$arcsec centred at $(0,0)$. The arrows show the effective deflection angle $\delta \ALP_{\textrm{eff}}$ due to another subdominant SIS with $\theta_{\textrm{Ep}}=0.1\,$arcsec at $(1'',0)$. The colors in the middle panel show the amplitudes in arcsec. }
\label{fig:SIS2-shift}
\end{figure*}

\begin{figure*} \center
\vspace{-2cm}
\includegraphics[width=15cm,pagebox=cropbox,clip]{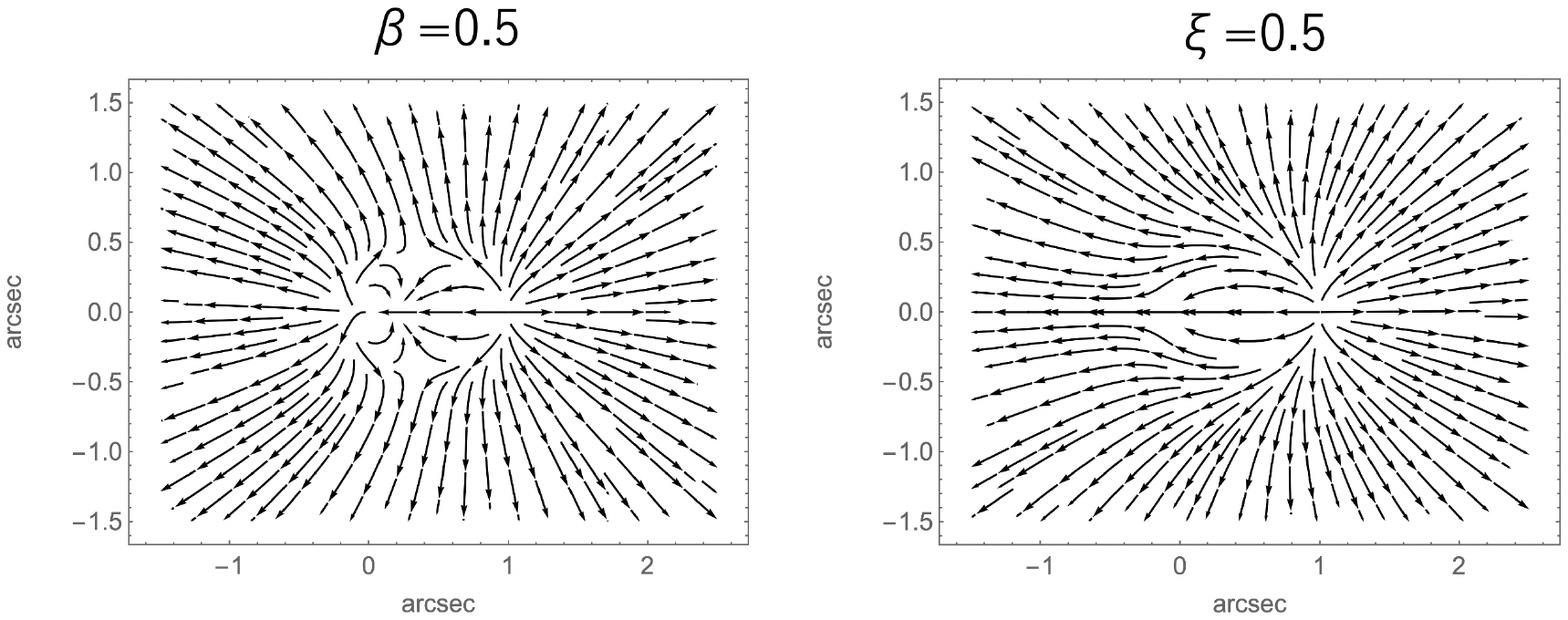}
\vspace{-2cm}
\caption{Streamlines of the effective deflection angles for an SIS. The model parameters are the same as in Fig. \ref{fig:SIS2-shift}. }
\label{fig:SIS2-shift-stream}
\end{figure*}

\begin{figure*} \center
\hspace{0.8cm}
\includegraphics[width=15.5cm,pagebox=cropbox,clip]{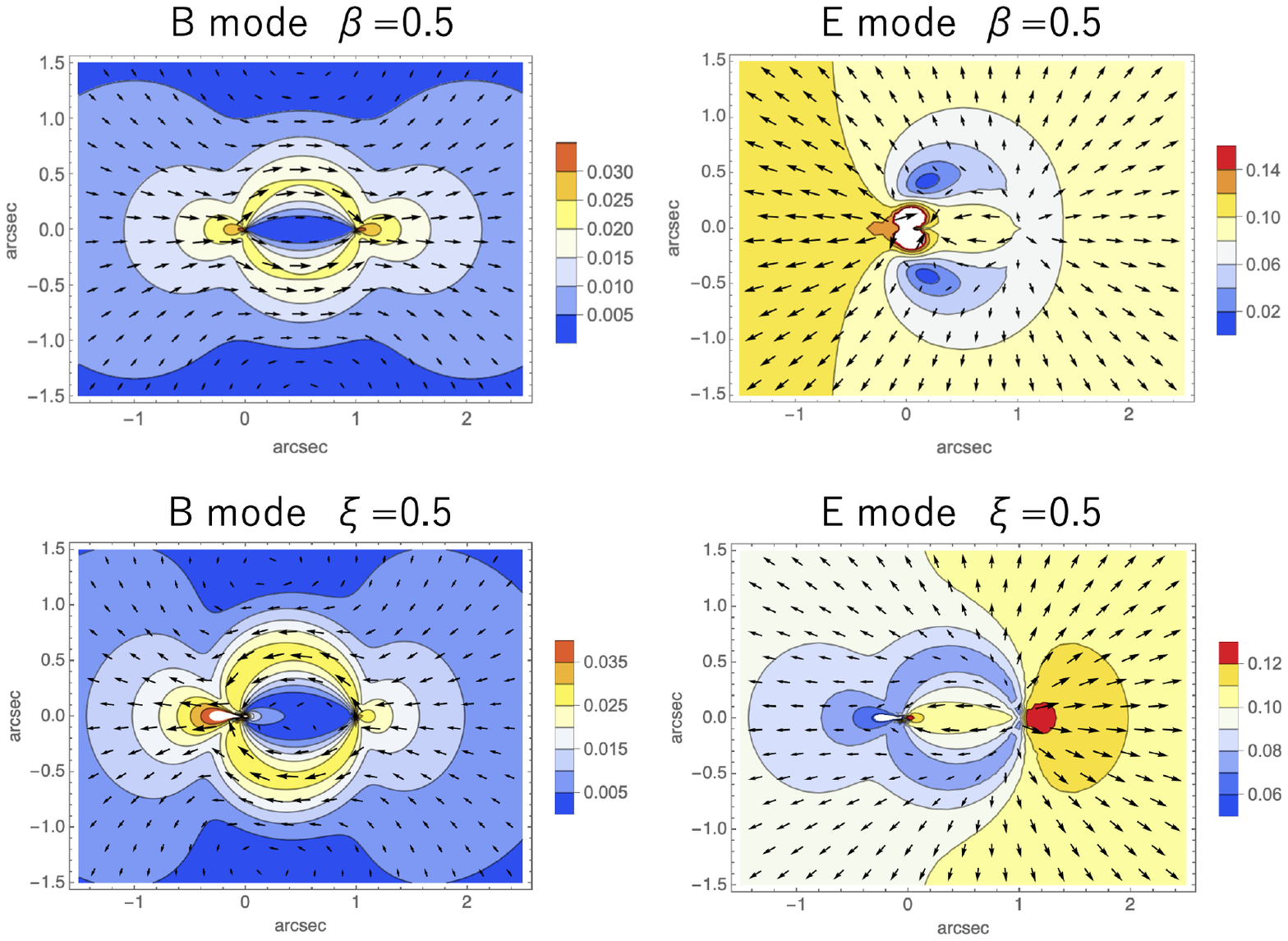}
\vspace{0.cm}
\caption{Astrometric lensing B and E-modes for an SIS. The arrows show the B and E components in the effective deflection angle  $\delta \ALP_{\textrm{eff}}$. The model parameters are the same as in Fig. \ref{fig:SIS2-shift}. The colors show the amplitudes of the shift in arcsec. }
\label{fig:SIS2-shift-EB}
\end{figure*}

The effective deflection angles $\delta \ALP_{\textrm{eff}}$ in an SIS+SIS model with $\theta_E=1\,$arcsec and $\theta_{E0}=0.1\,$arcsec are shown in Fig. \ref{fig:SIS2-shift}. If a subdominant SIS resides in the background of a dominant lens, the amplitude of the effective deflection angle $\delta \ALP_{\textrm{eff}}$ is constant but the direction is anisotropic. The directions resemble those of an electric field of a dipole. If a subdominant SIS resides in the foreground of a dominant lens, the amplitude of the effective deflection angle $\delta \ALP_{\textrm{eff}}$ is not constant and the direction is anisotropic. The streamlines resemble those in an SIS model with a background SIS with the same Einstein radius except for those in the central region of the dominant SIS (Fig. \ref{fig:SIS2-shift-stream}). For $\beta=0.5$ and $\xi=0.5$, the maximum amplitudes of the astrometric lensing B-mode $\delta \ALP_{\textrm{eff}}^\textrm{B}$ are $\sim 0.03\,$arcsec. As shown in Fig. \ref{fig:SIS2-shift-EB}, the contours of the amplitude of lensing B-mode have a dumbbell-like shape with a spindle-shaped void for both the foreground and background cases. In contrast, the contours of the amplitude of lensing E-mode $\delta \ALP_{\textrm{eff}}^\textrm{B}$ have a complex structure that depends on the position of the subdominant SIS. The amplitude is largest in the central region of the dominant SIS in the foreground case whereas the amplitude is largest in the central region of the subdominant SIS. This implies that any fitting without considering the difference between the subdominant lens plane and the dominant lens plane may lead to systematic residual in the positions of quadruple images of a point-like source.        

\begin{figure*} \center
\vspace{-2cm}
\hspace{0.5cm}
\includegraphics[width=15.5cm,pagebox=cropbox,clip]{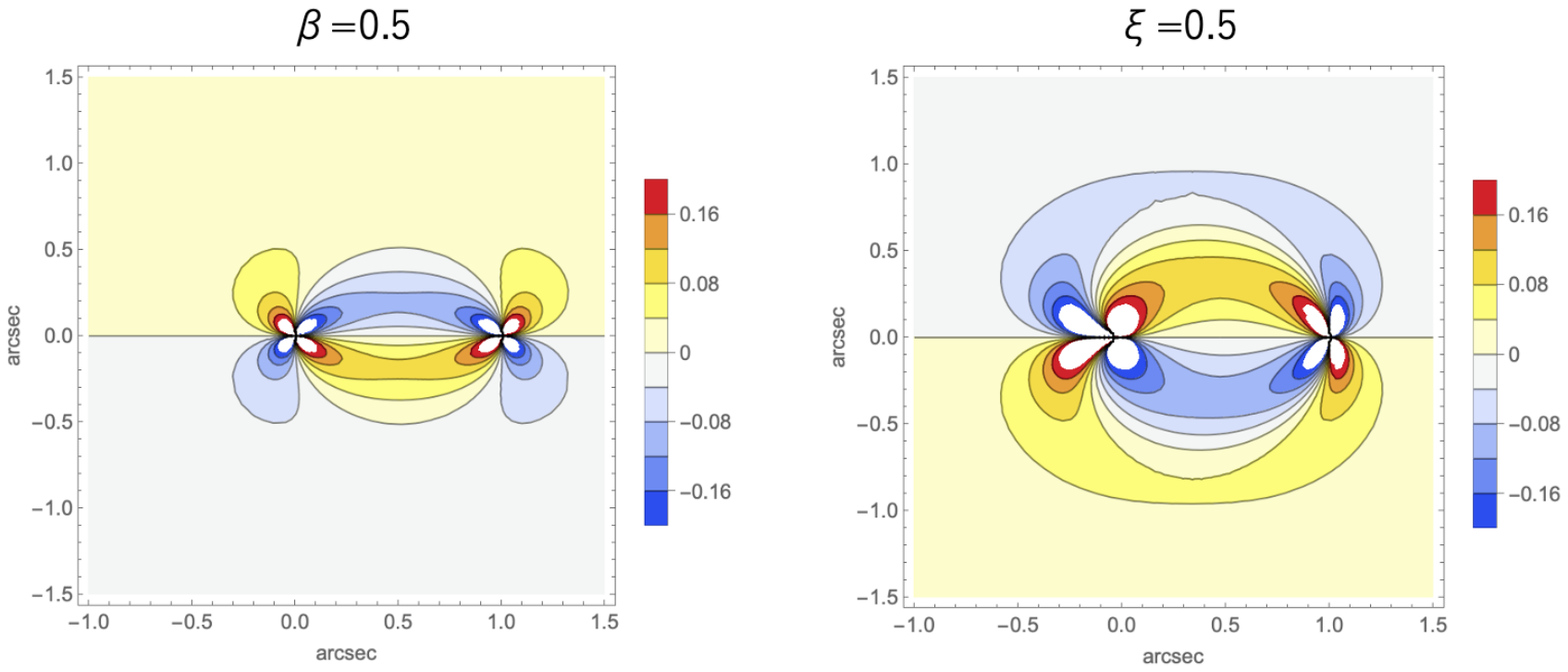}
\vspace{-2.5cm}
\caption{Rotation of effective deflection angle for an SIS with an Einstein radius of $\theta_\textrm{E0}=0.1\,$arcsec centred at $(1'',0)$. The colors show the amplitudes of rotation. The model parameters are the same as in Fig. \ref{fig:SIS2-shift}.  }
\label{fig:SIS2-rot}
\end{figure*}

As shown in Fig. \ref{fig:SIS2-rot}, the rotation of the effective deflection angle shows an octopole pattern that consists of a pair of quadrupoles centred at the centres of the SISs. The amplitudes of rotations are maximised in the diagonal lines and minimised in the vertical and horizontal directions of each SIS. The 'directions' of a rotation in the foreground SIS is opposite to that in the background SIS.     

\begin{figure*} \center
\vspace{-3cm}
\hspace{-0.5cm}
\includegraphics[width=16cm,pagebox=cropbox,clip]{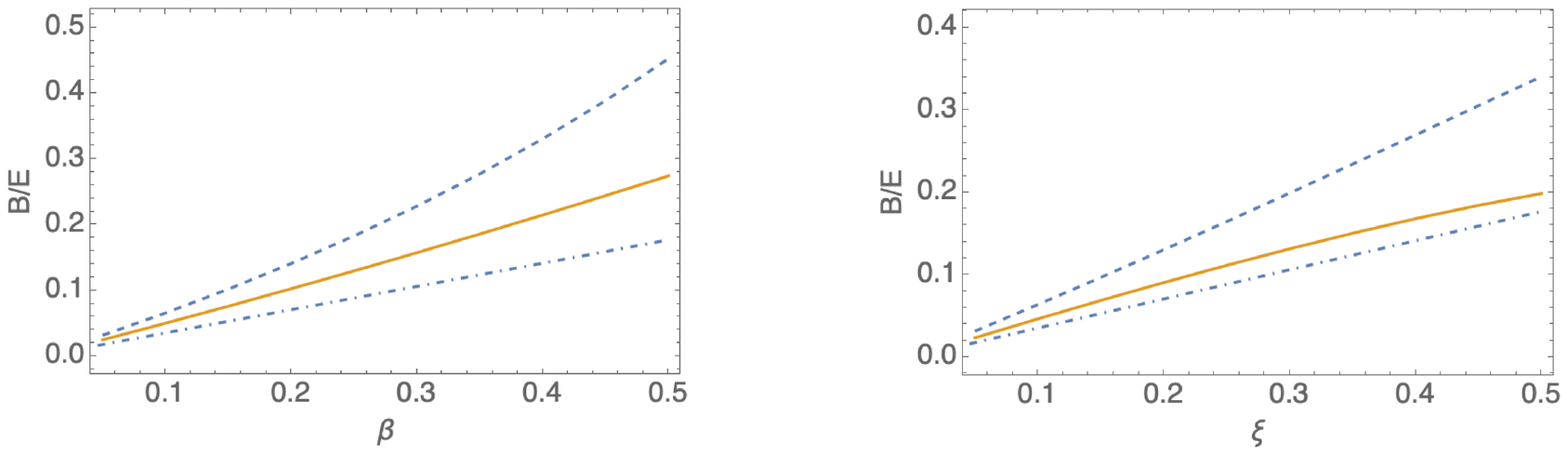}
\vspace{-3.5cm}
\caption{The ratio of B and E-modes for an SIS with an Einstein radius of $\theta_\textrm{E0}=0.1\,$arcsec centred at $(1'',0)$. The orange full curves denote the ratios of the averaged amplitude of B-mode astrometric shift and that of E-mode astrometric shift as a function of $\beta$ (top) and $\xi$ (bottom). Both the modes are averaged over an arc with $\theta=1.1\,$arcsec and $\phi=0.1\,$arcsec. The blue dashed curves show the ratio of averaged rotation and averaged convergence. Both the two contributions are averaged over the same arc. The blue dot-dashed curves represent the approximations of the ratios for the foreground case $\sqrt{2}\beta/4$ (left) and the background case $\sqrt{2}\xi/4$ (right). The model parameters are the same as in Fig. \ref{fig:SIS2-shift}.  }
\label{fig:SIS2-beta-approx-ave1}
\end{figure*}

\begin{figure*} \center
\hspace{-0.5cm}
\vspace{-3cm}
\includegraphics[width=15.5cm,pagebox=cropbox,clip]{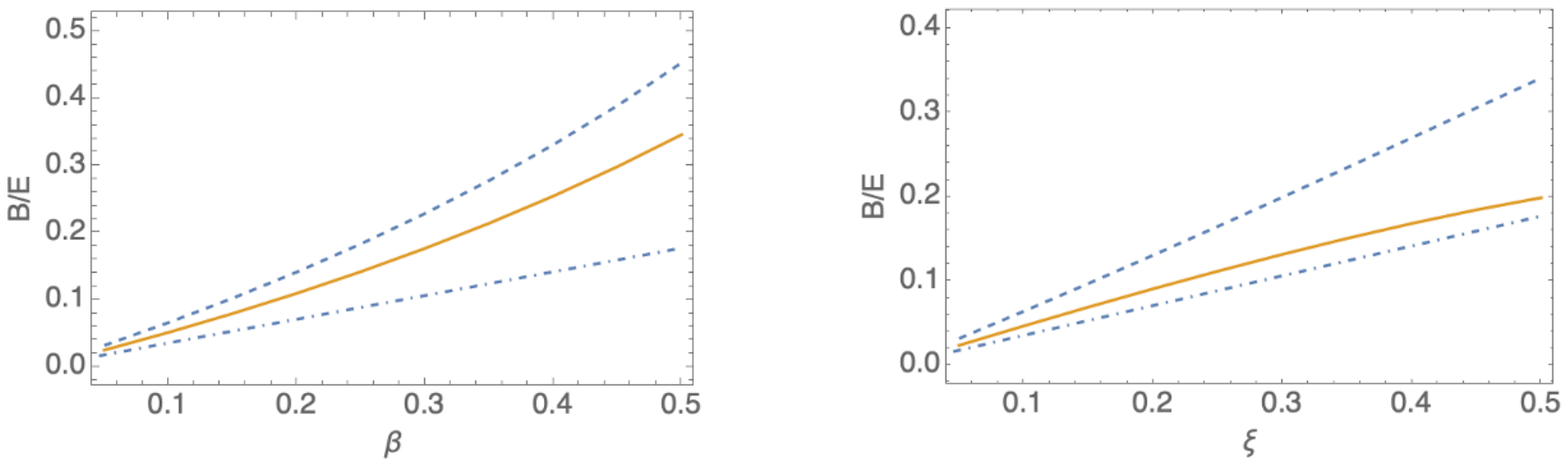}
\caption{The ratio of B and E-modes for an SIS with an Einstein radius of $\theta_\textrm{Ep}=0.03\,$arcsec centred at $(1'',0)$. The lines are the same in figure
\ref{fig:SIS2-beta-approx-ave1}. The model parameters except for $\theta_\textrm{Ep}$ are the same as in Fig. \ref{fig:SIS2-shift}.   }
\label{fig:SIS2-beta-approx-ave2}
\end{figure*}

In Fig. \ref{fig:SIS2-beta-approx-ave1} and Fig. \ref{fig:SIS2-beta-approx-ave2}, we show the ratio of the B-mode to E-mode $r^{\textrm {BE}}=|\delta \ALP_{\textrm{eff}}^{\textrm{B}} |/|\delta \ALP_{\textrm{eff}}^{\textrm{E}}|$, averaged over an arc with $\theta=1.1\,$arcsec and $\phi=0.1\,$arcsec. We did not use rotations in the Einstein ring of the dominant lens for averaging because the rotations in such regions are extremely small especially in the vicinity of the centre of the subdominant SIS. The ratio $r^{\textrm{BE}}$ is a monotonically increasing function of $\beta$ or $\xi$, which equals $\sim 0.3$ for $\beta=0.5$, and $\sim 0.2$ for $\xi=0.5$. We found that the ratio of rotation to divergence, $|\nabla \times \delta \ALP_{\textrm{eff}}^{\textrm{B}} |/|\nabla \cdot \delta \ALP_{\textrm{eff}}^{\textrm{E}}|$ averaged over the arc are larger than $r^\textrm{BE}$. The analytic formula in eq. (\ref{eq:etaest2}) gives a good estimate up to moderate values of $\beta \sim 0.5$ and $\xi \sim 0.5$. 

\begin{figure*} \center
\hspace{0.5cm}
\includegraphics[width=15.5cm,pagebox=cropbox,clip]{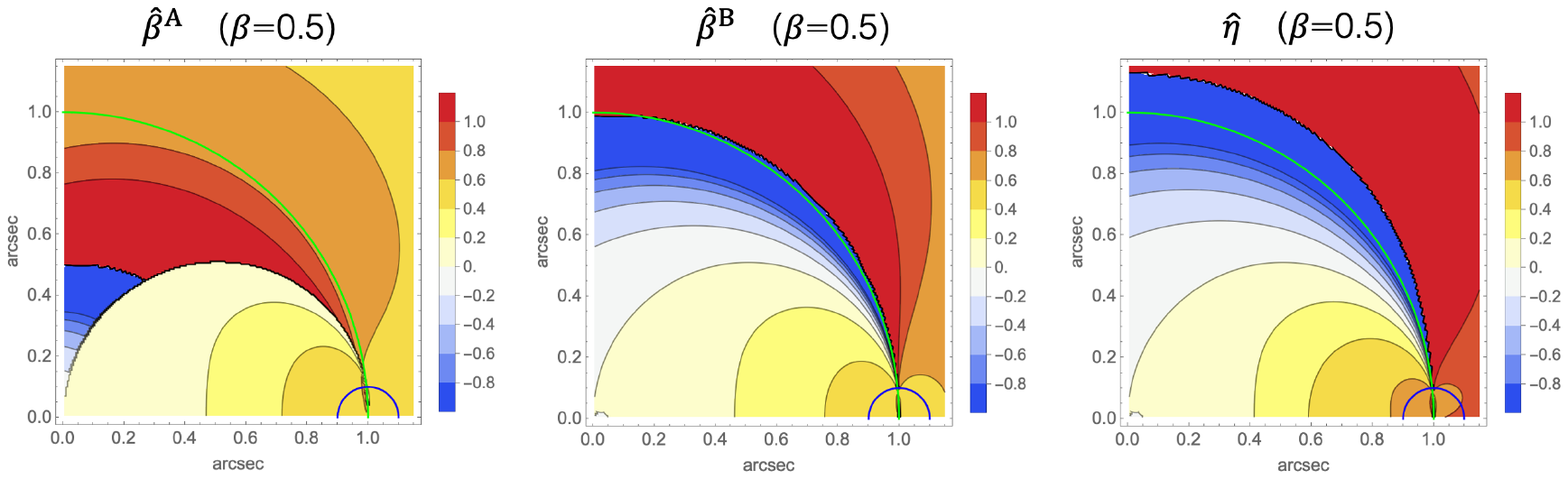}
\vspace{-3.5cm}
\caption{Approximated distance ratios in the lens plane for an SIS with $\beta=0.5$. The colors show the approximated distance ratios for $\hat{\beta}^{\textrm{A}}$ (left), 
$\hat{\beta}^{\textrm{B}}$ (middle), and $\hat{\eta}$ (right). Green curves denote the Einstein radius $\theta_\textrm{E}$ of the dominant SIS and the blue curves represent the Einstein radius $\theta_\textrm{Ep}$ of the subdominant SIS around the centre. The model parameters are the same as in Fig. \ref{fig:SIS2-shift}.}
  
\label{fig:SIS2-beta-approx}
\end{figure*}

Next, we show the values of the distance ratio estimators $\hat{\beta}^{\textrm{A}}$, $\hat{\beta}^{\textrm{B}}$, and $\hat{\eta}$ in the dominant lens plane in Fig. \ref{fig:SIS2-beta-approx}. We found that these estimators give a good approximation in the vicinity of the subdominant SIS. $\hat{\beta}^{\textrm{B}}$ and $\hat{\eta}$ give a worse result around the Einstein ring of the dominant lens in which the rotation is almost zero. For $\beta=0.5$, $\hat{\beta}^{\textrm{A}}$ gives the best result. 

\begin{figure*} \center
\hspace{1.5cm}
\includegraphics[width=15.5cm,pagebox=cropbox,clip]{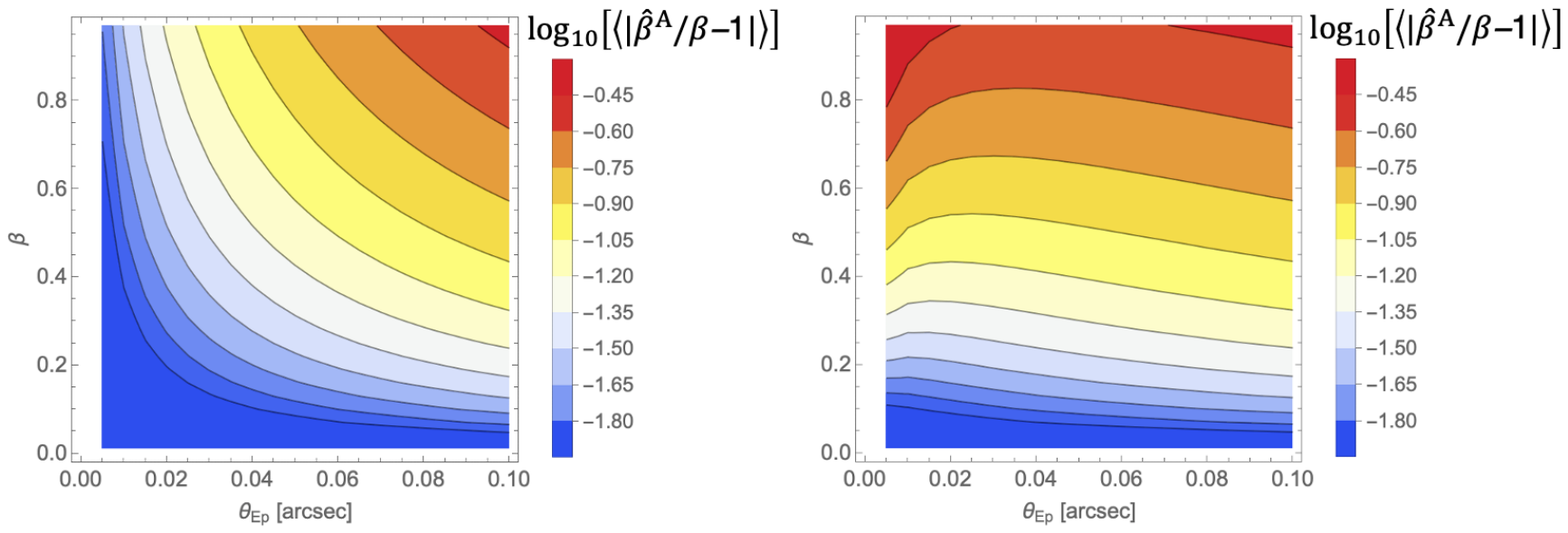}
\vspace{-3cm}
\caption{The effect of sample region for foreground SIS perturbers. The colors show the common logarithm of mean relative errors of the distance ratio estimator $\hat{\beta}^\textrm{A}$ as a function of $\beta$ and $\theta_\textrm{Ep}$. The center of a subdominant SIS is fixed to $(x_1,x_2)=(1'',0)$. The mean is obtained from averaging the relative errors on an arc with a radius of $1''+\theta_\textrm{Ep}$ that subtends an azimuthal angle of $\phi=\theta_\textrm{Ep}$ (left) and those on an arc with a radius of $1.1\,$arcsec that subtends an azimuthal angle of $\phi = 0.1$\,rad (right). }

\label{fig:SIS2-beta-approx-gamma}
\end{figure*}

Finally, we show the effect of sample region, which was used to estimate the distant ratio using $\hat{\beta}^{\textrm{A}}$ in Fig. \ref{fig:SIS2-beta-approx-gamma}. The center of a subdominant SIS is fixed to $(x_1,x_2)=(1'',0)$. We consider two types of sample regions: 1) an arc with a radius of $1''+\theta_\textrm{Ep}$ that subtends an azimuthal angle of $\phi=\theta_\textrm{Ep}$, 2) an arc with a radius of $1.1\,$arcsec that subtends an azimuthal angle of $\phi = 0.1$\,rad. Note that we avoided the neighbourhood of an Einstein ring of the dominant lens as the rotation is very small. The errors in the case 1) is much smaller than the case 2) as the sample region is much closer to the centre of the subdominant SIS. For $\beta<0.5$, the relative errors in both 1) and 2) were found to be $\lesssim 0.1$. Even for large distance ratios $\beta>0.5$, the relative errors are $<0.1$ for $\theta_\textrm{Ep}<0.04\,$arcsec in case 1). Thus, the foreground distance ratio estimator $\hat{\beta}^{\textrm{A}}$ gives a relatively accurate approximation if it is used for estimating the property of the central region of a perturbing halo. Even if the lensed arc of an extended source is not on the centre of a perturbing halo, $\hat{\beta}^{\textrm{A}}$ can be used as a good estimator if $\beta$ is sufficiently small. 

\section{Conclusion and Discussion}

In this study, we investigated the characteristics of astrometric lensing B-mode in strong lensing systems that consist of a dominant and a subdominant lens residing at distinct redshifts. The B-mode arises from the coupling between strong lensing induced by a dominant lens, and weak lensing generated by a subdominant lens. By measuring both B and E-modes, we can deduce the distance ratio, and 'bare' convergence and shear perturbations attributed to the subdominant lens.

In cases where a subdominant lens is located behind the dominant lens, we can derive an exact formula if the dominant lens is perfectly modelled. However, when the situation reverses, with a dominant lens resides behind a subdominant lens, we cannot obtain an exact formula even with a perfect model of the dominant lens. In such cases, we employ certain approximations that yield an exact value when the distance between the dominant and subdominant lenses approaches zero.

We demonstrated that any scale transformation in the distance ratio of a subdominant lens corresponds to a mass-sheet transformation in the background lens plane. Consequently, determining the distance ratio necessitates assumptions about the values of the mass-sheet within the background lens plane. Nevertheless, if we measure time delays between perturbed multiple lensed images and know the redshifts of a subdominant lens, we can break the mass-sheet degeneracy for a given $H_0$, enabling us to determine it without uncertainty. Moreover, if the redshifts of both foreground and background perturbers, as well as the dominant lens and the source, are known, observations of astrometric lensing B-mode can break the mass-sheet degeneracies. This would lead to a reduction in systematic errors in the estimated value of $H_0$. 

Our analysis focuses on systems with a single subdominant lens whose deflection angle is significantly smaller than that of the dominant lens. In reality, the gravitational influence of multiple subdominant lenses along each photon's path must be considered. When both the second and third dominant lenses are located either in the foreground or the background of the dominant lens, the rotation signal may be amplified. Conversely, if they are positioned separately in the foreground and background, the signal could weaken due to cancellation. Investigating these effects in systems comprising three or more lenses falls beyond the scope of this paper.

Assuming that the impact of model degeneracy due to the extended multi-plane mass-sheet transformation (eMMST) we discussed is negligible, the measurement of astrometric lensing B-modes holds the potential to constrain the abundance of intergalactic dark haloes with masses of $\lesssim 10^8\,\ms$ along the line of sight (LOS) in quasar-galaxy or galaxy-galaxy strong lens systems.

To measure astrometric shifts in gravitational lensing systems, we adopt the following framework:

\begin{itemize}
\item \textbf{Lens System:} We consider a system dominated by a single lens with a smooth gravitational potential.  Subdominant lenses may be present, but their gravitational potential fluctuates on scales comparable to the Einstein radius $\theta_\mathrm{E}$ of the dominant lens. This allows us to separate the effects of the dominant lens from those of smaller perturbers.

\item \textbf{Source Structure:} The source is modeled as a combination of a point-like component (e.g., quasar core emission) and an extended component (e.g., thermal dust emission or non-thermal jet emission). The angular scale of intensity fluctuations in the source $\theta_\mathrm{S}$ satisfies the following condition:
    \begin{equation*}
    \delta \alpha \ll \theta_\mathrm{S} \ll \theta_\mathrm{E},
    \end{equation*}
    where $\delta \alpha$ is the scale of the astrometric shifts. This hierarchy of scales is crucial for our analysis.

\item \textbf{Astrometric Shift Scale:} The angular scale of fluctuations in the astrometric shift $\theta_\mathrm{A}$ is assumed to be comparable to or slightly smaller than the Einstein radius: $\theta_\mathrm{A} \lesssim \theta_\mathrm{E}$. This assumption is justified by the fact that the dominant lens primarily determines the overall lensing geometry.
\end{itemize}
These assumptions are consistent with observations of typical quadruply lensed quasars, such as MG\,J0414+0534 in the submillimeter band, where $\delta \alpha \sim 0.005\,$arcsec, $\theta_\mathrm{S} = 0.01\,$arcsec - $0.1\,$arcsec, and $\theta_\mathrm{E} \sim \theta_\textrm{A} \sim 1\,$arcsec \cite{inoue2023}.

We utilize the following procedure to reconstruct the deflection field:
\begin{enumerate}
\item \textbf{Averaged Source Image:}  An 'averaged' source image is constructed using the positions and fluxes of the point-like images, assuming a smooth lens potential without perturbations \cite{inoue2016,inoue2023}. This provides a baseline image that accounts for the dominant lensing effects.

\item \textbf{Deflection Field Estimation:}  The deflection field is obtained by comparing the 'averaged' source image with the de-lensed observed images. The difference between these images reveals the perturbations caused by subdominant lenses.

\item \textbf{Error Mitigation:} The 'averaged' image may deviate slightly from the true de-lensed image. However, this difference is expected to be much smaller than $\delta \alpha$ because multiple images (typically four) are used to construct the 'averaged' image. Furthermore, since $\delta \alpha \ll \theta_\mathrm{S} \ll \theta_\mathrm{A}$, we can employ numerous random source samples to minimize the residual error.
\end{enumerate}
It is important to note that any local translation in the astrometric shifts within the source plane is unobservable (see footnote in Section 2.2). Nevertheless, we anticipate that such specific configurations of gravitational perturbations are negligible, as line-of-sight perturbers are unlikely to be correlated with the dominant lens.

Our toy models suggest that feasibility of measuring B-modes in astrometric shifts can be assessed as follows: As a reference lens model, we employ a dominant SIS with an Einstein radius of $1\,$arcsec. Then, a subdominant SIS with an Einstein radius of $\gtrsim 0.03\,$arcsec in the LOS would produce B-modes with a shift of $\gtrsim 0.003\,$arcsec for $\beta>0.2$ or $\xi>0.2$. These shifts can be observed with telescopes featuring an angular resolution of $\lesssim 0.03\,$arcsec, assuming a typical magnification of $\mu\sim 10$ for lensed image separations of $\lesssim 0.5\,$arcsec \cite{inoue2005b}. Hence, instruments like the Atacama Large Millimeter/Submillimeter Array (ALMA) possess the capability to detect astrometric lensing B-modes resulting from less massive LOS haloes, provided that the distance between the dominant and subdominant lenses is sufficiently large and the signal-to-noise ratio of intensity in the lens plane is suitably high.

In the near future, we plan to investigate the practicality of measuring astrometric lensing B-modes with ALMA and other instruments such as the next generation Very Large Array (ngVLA)\cite{ngVLA2018}, utilizing more sophisticated models and taking into account observational capabilities.

% \section{Acknowledgements}
\section{Data Availability}
This theoretical paper does not contain empirical data; rather, it presents conceptual frameworks, mathematical models, and computational simulations to support its findings and conclusions. The data relevant to this paper's reproducibility and validation are primarily comprised of computer codes, mathematical equations, and simulation parameters. For inquiries related to data availability, questions about the theoretical framework, or assistance with replicating the results, please contact the author.

\bibliographystyle{JHEP}
\bibliography{2023BmodeLOS-JCAP}

\providecommand{\href}[2]{#2}\begingroup\raggedright\begin{thebibliography}{10}

\bibitem{kauffmann1993}
G.~{Kauffmann}, S.D.M.~{White} and B.~{Guiderdoni}, \emph{{The formation and
  evolution of galaxies within merging dark matter haloes.}},
  \href{https://doi.org/10.1093/mnras/264.1.201}{\emph{Mon. Not. R. Astron.
  Soc.} {\bfseries 264} (1993) 201}.

\bibitem{klypin1999}
A.~{Klypin}, A.V.~{Kravtsov}, O.~{Valenzuela} and F.~{Prada}, \emph{{Where Are
  the Missing Galactic Satellites?}},
  \href{https://doi.org/10.1086/307643}{\emph{\apj} {\bfseries 522} (1999) 82}
  [\href{https://arxiv.org/abs/astro-ph/9901240}{{\ttfamily
  astro-ph/9901240}}].

\bibitem{moore1999}
B.~{Moore}, S.~{Ghigna}, F.~{Governato}, G.~{Lake}, T.~{Quinn}, J.~{Stadel}
  et~al., \emph{{Dark Matter Substructure within Galactic Halos}},
  \href{https://doi.org/10.1086/312287}{\emph{\apj} {\bfseries 524} (1999) L19}
  [\href{https://arxiv.org/abs/astro-ph/9907411}{{\ttfamily
  astro-ph/9907411}}].

\bibitem{wetzel2016}
A.R.~{Wetzel}, P.F.~{Hopkins}, J.-h.~{Kim}, C.-A.~{Faucher-Gigu{\`e}re},
  D.~{Keres} and E.~{Quataert}, \emph{{Reconciling Dwarf Galaxies with
  {\ensuremath{\Lambda}}CDM Cosmology: Simulating a Realistic Population of
  Satellites around a Milky Way-mass Galaxy}},
  \href{https://doi.org/10.3847/2041-8205/827/2/L23}{\emph{\apj} {\bfseries
  827} (2016) L23} [\href{https://arxiv.org/abs/1602.05957}{{\ttfamily
  1602.05957}}].

\bibitem{brooks2017}
A.M.~{Brooks}, E.~{Papastergis}, C.R.~{Christensen}, F.~{Governato},
  A.~{Stilp}, T.R.~{Quinn} et~al., \emph{{How to Reconcile the Observed
  Velocity Function of Galaxies with Theory}},
  \href{https://doi.org/10.3847/1538-4357/aa9576}{\emph{\apj} {\bfseries 850}
  (2017) 97} [\href{https://arxiv.org/abs/1701.07835}{{\ttfamily 1701.07835}}].

\bibitem{fielder2019}
C.E.~{Fielder}, Y.-Y.~{Mao}, J.A.~{Newman}, A.R.~{Zentner} and T.C.~{Licquia},
  \emph{{Predictably missing satellites: subhalo abundances in Milky Way-like
  haloes}}, \href{https://doi.org/10.1093/mnras/stz1098}{\emph{Mon. Not. R.
  Astron. Soc.} {\bfseries 486} (2019) 4545}
  [\href{https://arxiv.org/abs/1807.05180}{{\ttfamily 1807.05180}}].

\bibitem{nashimoto2022}
M.~{Nashimoto}, M.~{Tanaka}, M.~{Chiba}, K.~{Hayashi}, Y.~{Komiyama} and
  T.~{Okamoto}, \emph{{The Missing Satellite Problem outside of the Local
  Group. II. Statistical Properties of Satellites of Milky Way-like Galaxies}},
  \href{https://doi.org/10.3847/1538-4357/ac83a4}{\emph{\apj} {\bfseries 936}
  (2022) 38} [\href{https://arxiv.org/abs/2207.11992}{{\ttfamily 2207.11992}}].

\bibitem{mao1998}
S.~{Mao} and P.~{Schneider}, \emph{{Evidence for substructure in lens
  galaxies?}},
  \href{https://doi.org/10.1046/j.1365-8711.1998.01319.x}{\emph{Mon. Not. R.
  Astron. Soc.} {\bfseries 295} (1998) 587}
  [\href{https://arxiv.org/abs/astro-ph/9707187}{{\ttfamily
  astro-ph/9707187}}].

\bibitem{metcalf2001}
R.B.~{Metcalf} and P.~{Madau}, \emph{{Compound Gravitational Lensing as a Probe
  of Dark Matter Substructure within Galaxy Halos}},
  \href{https://doi.org/10.1086/323695}{\emph{\apj} {\bfseries 563} (2001) 9}
  [\href{https://arxiv.org/abs/astro-ph/0108224}{{\ttfamily
  astro-ph/0108224}}].

\bibitem{chiba2002}
M.~{Chiba}, \emph{{Probing Dark Matter Substructure in Lens Galaxies}},
  \href{https://doi.org/10.1086/324493}{\emph{\apj} {\bfseries 565} (2002) 17}
  [\href{https://arxiv.org/abs/astro-ph/0109499}{{\ttfamily
  astro-ph/0109499}}].

\bibitem{dalal-kochanek2002}
N.~{Dalal} and C.S.~{Kochanek}, \emph{{Direct Detection of Cold Dark Matter
  Substructure}}, \href{https://doi.org/10.1086/340303}{\emph{\apj} {\bfseries
  572} (2002) 25} [\href{https://arxiv.org/abs/astro-ph/0111456}{{\ttfamily
  astro-ph/0111456}}].

\bibitem{keeton2003}
C.R.~Keeton, B.S.~Gaudi and A.O.~Petters, \emph{Identifying lenses with
  small-scale structure. i. cusp lenses},
  \href{https://doi.org/10.1086/378934}{\emph{\apj} {\bfseries 598} (2003)
  138}.

\bibitem{inoue-chiba2003}
K.T.~{Inoue} and M.~{Chiba}, \emph{{Direct Mapping of Massive Compact Objects
  in Extragalactic Dark Halos}},
  \href{https://doi.org/10.1086/377247}{\emph{\apj} {\bfseries 591} (2003) L83}
  [\href{https://arxiv.org/abs/astro-ph/0304474}{{\ttfamily
  astro-ph/0304474}}].

\bibitem{inoue2005a}
K.T.~{Inoue} and M.~{Chiba}, \emph{{Extended Source Effects in Substructure
  Lensing}}, \href{https://doi.org/10.1086/496870}{\emph{\apj} {\bfseries 634}
  (2005) 77} [\href{https://arxiv.org/abs/astro-ph/0411168}{{\ttfamily
  astro-ph/0411168}}].

\bibitem{xu2009}
D.D.~{Xu}, S.~{Mao}, J.~{Wang}, V.~{Springel}, L.~{Gao}, S.D.M.~{White} et~al.,
  \emph{{Effects of dark matter substructures on gravitational lensing: results
  from the Aquarius simulations}},
  \href{https://doi.org/10.1111/j.1365-2966.2009.15230.x}{\emph{Mon. Not. R.
  Astron. Soc.} {\bfseries 398} (2009) 1235}
  [\href{https://arxiv.org/abs/0903.4559}{{\ttfamily 0903.4559}}].

\bibitem{xu2010}
D.D.~{Xu}, S.~{Mao}, A.P.~{Cooper}, J.~{Wang}, L.~{Gao}, C.S.~{Frenk} et~al.,
  \emph{{Substructure lensing: effects of galaxies, globular clusters and
  satellite streams}},
  \href{https://doi.org/10.1111/j.1365-2966.2010.17235.x}{\emph{Mon. Not. R.
  Astron. Soc.} {\bfseries 408} (2010) 1721}
  [\href{https://arxiv.org/abs/1004.3094}{{\ttfamily 1004.3094}}].

\bibitem{evans2003}
N.W.~{Evans} and H.J.~{Witt}, \emph{{Fitting gravitational lenses: truth or
  delusion}},
  \href{https://doi.org/10.1046/j.1365-2966.2003.07057.x}{\emph{Mon. Not. R.
  Astron. Soc.} {\bfseries 345} (2003) 1351}
  [\href{https://arxiv.org/abs/astro-ph/0212013}{{\ttfamily
  astro-ph/0212013}}].

\bibitem{oguri2005}
M.~{Oguri}, \emph{{Lens galaxy environments and anomalous flux ratios in
  gravitational lenses}},
  \href{https://doi.org/10.1111/j.1745-3933.2005.00061.x}{\emph{Mon. Not. R.
  Astron. Soc.} {\bfseries 361} (2005) L38}
  [\href{https://arxiv.org/abs/astro-ph/0411464}{{\ttfamily
  astro-ph/0411464}}].

\bibitem{gilman2017}
D.~{Gilman}, A.~{Agnello}, T.~{Treu}, C.R.~{Keeton} and A.M.~{Nierenberg},
  \emph{{Strong lensing signatures of luminous structure and substructure in
  early-type galaxies}}, \href{https://doi.org/10.1093/mnras/stx158}{\emph{Mon.
  Not. R. Astron. Soc.} {\bfseries 467} (2017) 3970}
  [\href{https://arxiv.org/abs/1610.08525}{{\ttfamily 1610.08525}}].

\bibitem{hsueh2017}
J.W.~{Hsueh}, L.~{Oldham}, C.~{Spingola}, S.~{Vegetti}, C.D.~{Fassnacht},
  M.W.~{Auger} et~al., \emph{{SHARP - IV. An apparent flux-ratio anomaly
  resolved by the edge-on disc in B0712+472}},
  \href{https://doi.org/10.1093/mnras/stx1082}{\emph{Mon. Not. R. Astron. Soc.}
  {\bfseries 469} (2017) 3713}
  [\href{https://arxiv.org/abs/1701.06575}{{\ttfamily 1701.06575}}].

\bibitem{hsueh2018}
J.-W.~{Hsueh}, G.~{Despali}, S.~{Vegetti}, D.~{Xu}, C.D.~{Fassnacht} and
  R.B.~{Metcalf}, \emph{{Flux-ratio anomalies from discs and other baryonic
  structures in the Illustris simulation}},
  \href{https://doi.org/10.1093/mnras/stx3320}{\emph{Mon. Not. R. Astron. Soc.}
  {\bfseries 475} (2018) 2438}
  [\href{https://arxiv.org/abs/1707.07680}{{\ttfamily 1707.07680}}].

\bibitem{treu-koopmans2004}
T.~{Treu} and L.V.E.~{Koopmans}, \emph{{Massive Dark Matter Halos and Evolution
  of Early-Type Galaxies to z \raisebox{-0.5ex}\textasciitilde 1}},
  \href{https://doi.org/10.1086/422245}{\emph{\apj} {\bfseries 611} (2004) 739}
  [\href{https://arxiv.org/abs/astro-ph/0401373}{{\ttfamily
  astro-ph/0401373}}].

\bibitem{koopmans2005}
L.V.E.~{Koopmans}, \emph{{Gravitational imaging of cold dark matter
  substructures}},
  \href{https://doi.org/10.1111/j.1365-2966.2005.09523.x}{\emph{Mon. Not. R.
  Astron. Soc.} {\bfseries 363} (2005) 1136}
  [\href{https://arxiv.org/abs/astro-ph/0501324}{{\ttfamily
  astro-ph/0501324}}].

\bibitem{vegetti2009}
S.~{Vegetti} and L.V.E.~{Koopmans}, \emph{{Bayesian strong gravitational-lens
  modelling on adaptive grids: objective detection of mass substructure in
  Galaxies}},
  \href{https://doi.org/10.1111/j.1365-2966.2008.14005.x}{\emph{Mon. Not. R.
  Astron. Soc.} {\bfseries 392} (2009) 945}
  [\href{https://arxiv.org/abs/0805.0201}{{\ttfamily 0805.0201}}].

\bibitem{vegetti2010}
S.~{Vegetti}, L.V.E.~{Koopmans}, A.~{Bolton}, T.~{Treu} and R.~{Gavazzi},
  \emph{{Detection of a dark substructure through gravitational imaging}},
  \href{https://doi.org/10.1111/j.1365-2966.2010.16865.x}{\emph{Mon. Not. R.
  Astron. Soc.} {\bfseries 408} (2010) 1969}
  [\href{https://arxiv.org/abs/0910.0760}{{\ttfamily 0910.0760}}].

\bibitem{chantry2010}
V.~{Chantry}, D.~{Sluse} and P.~{Magain}, \emph{{COSMOGRAIL: the COSmological
  MOnitoring of GRAvItational Lenses. VIII. Deconvolution of high resolution
  near-IR images and simple mass models for 7 gravitationally lensed quasars}},
  \href{https://doi.org/10.1051/0004-6361/200912971}{\emph{Astron. \&
  Astrophys.} {\bfseries 522} (2010) A95}
  [\href{https://arxiv.org/abs/1007.3142}{{\ttfamily 1007.3142}}].

\bibitem{vegetti2012}
S.~{Vegetti}, D.J.~{Lagattuta}, J.P.~{McKean}, M.W.~{Auger}, C.D.~{Fassnacht}
  and L.V.E.~{Koopmans}, \emph{{Gravitational detection of a low-mass dark
  satellite galaxy at cosmological distance}},
  \href{https://doi.org/10.1038/nature10669}{\emph{Nature} {\bfseries 481}
  (2012) 341} [\href{https://arxiv.org/abs/1201.3643}{{\ttfamily 1201.3643}}].

\bibitem{vegetti2014}
S.~{Vegetti}, L.V.E.~{Koopmans}, M.W.~{Auger}, T.~{Treu} and A.S.~{Bolton},
  \emph{{Inference of the cold dark matter substructure mass function at z =
  0.2 using strong gravitational lenses}},
  \href{https://doi.org/10.1093/mnras/stu943}{\emph{Mon. Not. R. Astron. Soc.}
  {\bfseries 442} (2014) 2017}
  [\href{https://arxiv.org/abs/1405.3666}{{\ttfamily 1405.3666}}].

\bibitem{inoue-minezaki2016}
K.T.~{Inoue}, T.~{Minezaki}, S.~{Matsushita} and M.~{Chiba}, \emph{{ALMA
  imprint of intergalactic dark structures in the gravitational lens SDP.81}},
  \href{https://doi.org/10.1093/mnras/stw168}{\emph{Mon. Not. R. Astron. Soc.}
  {\bfseries 457} (2016) 2936}
  [\href{https://arxiv.org/abs/1510.00150}{{\ttfamily 1510.00150}}].

\bibitem{hezaveh2016a}
Y.D.~{Hezaveh}, N.~{Dalal}, D.P.~{Marrone}, Y.-Y.~{Mao}, W.~{Morningstar},
  D.~{Wen} et~al., \emph{{Detection of Lensing Substructure Using ALMA
  Observations of the Dusty Galaxy SDP.81}},
  \href{https://doi.org/10.3847/0004-637X/823/1/37}{\emph{\apj} {\bfseries 823}
  (2016) 37} [\href{https://arxiv.org/abs/1601.01388}{{\ttfamily 1601.01388}}].

\bibitem{chatterjee2018}
S.~{Chatterjee} and L.V.E.~{Koopmans}, \emph{{The inner mass power spectrum of
  galaxies using strong gravitational lensing: beyond linear approximation}},
  \href{https://doi.org/10.1093/mnras/stx2674}{\emph{Mon. Not. R. Astron. Soc.}
  {\bfseries 474} (2018) 1762}
  [\href{https://arxiv.org/abs/1710.03075}{{\ttfamily 1710.03075}}].

\bibitem{cagan2020}
A.~{{\c{C}}agan {\c{S}}eng{\"u}l}, A.~{Tsang}, A.~{Diaz Rivero}, C.~{Dvorkin},
  H.-M.~{Zhu} and U.~{Seljak}, \emph{{Quantifying the line-of-sight halo
  contribution to the dark matter convergence power spectrum from strong
  gravitational lenses}},
  \href{https://doi.org/10.1103/PhysRevD.102.063502}{\emph{\prd} {\bfseries
  102} (2020) 063502} [\href{https://arxiv.org/abs/2006.07383}{{\ttfamily
  2006.07383}}].

\bibitem{metcalf2005a}
R.B.~{Metcalf}, \emph{{The Importance of Intergalactic Structure to
  Gravitationally Lensed Quasars}},
  \href{https://doi.org/10.1086/431574}{\emph{\apj} {\bfseries 629} (2005) 673}
  [\href{https://arxiv.org/abs/astro-ph/0412538}{{\ttfamily
  astro-ph/0412538}}].

\bibitem{xu2012}
D.D.~{Xu}, S.~{Mao}, A.P.~{Cooper}, L.~{Gao}, C.S.~{Frenk}, R.E.~{Angulo}
  et~al., \emph{{On the effects of line-of-sight structures on lensing
  flux-ratio anomalies in a {$\Lambda$}CDM universe}},
  \href{https://doi.org/10.1111/j.1365-2966.2012.20484.x}{\emph{Mon. Not. R.
  Astron. Soc.} {\bfseries 421} (2012) 2553}
  [\href{https://arxiv.org/abs/1110.1185}{{\ttfamily 1110.1185}}].

\bibitem{inoue-takahashi2012}
K.T.~{Inoue} and R.~{Takahashi}, \emph{{Weak lensing by line-of-sight haloes as
  the origin of flux-ratio anomalies in quadruply lensed QSOs}},
  \href{https://doi.org/10.1111/j.1365-2966.2012.21915.x}{\emph{Mon. Not. R.
  Astron. Soc.} {\bfseries 426} (2012) 2978}
  [\href{https://arxiv.org/abs/1207.2139}{{\ttfamily 1207.2139}}].

\bibitem{takahashi-inoue2014}
R.~{Takahashi} and K.T.~{Inoue}, \emph{{Weak lensing by intergalactic
  ministructures in quadruple lens systems: simulation and detection}},
  \href{https://doi.org/10.1093/mnras/stu328}{\emph{Mon. Not. R. Astron. Soc.}
  {\bfseries 440} (2014) 870}
  [\href{https://arxiv.org/abs/1308.4855}{{\ttfamily 1308.4855}}].

\bibitem{inoue2016}
K.T.~{Inoue}, \emph{{On the origin of the flux ratio anomaly in quadruple lens
  systems}}, \href{https://doi.org/10.1093/mnras/stw1270}{\emph{Mon. Not. R.
  Astron. Soc.} {\bfseries 461} (2016) 164}
  [\href{https://arxiv.org/abs/1601.04414}{{\ttfamily 1601.04414}}].

\bibitem{despali2018}
G.~{Despali}, S.~{Vegetti}, S.D.M.~{White}, C.~{Giocoli} and F.C.~{van den
  Bosch}, \emph{{Modelling the line-of-sight contribution in substructure
  lensing}}, \href{https://doi.org/10.1093/mnras/sty159}{\emph{Mon. Not. R.
  Astron. Soc.} {\bfseries 475} (2018) 5424}
  [\href{https://arxiv.org/abs/1710.05029}{{\ttfamily 1710.05029}}].

\bibitem{sengul2022}
A.{\c{C}}.~{Seng{\"u}l}, C.~{Dvorkin}, B.~{Ostdiek} and A.~{Tsang},
  \emph{{Substructure detection reanalysed: dark perturber shown to be a
  line-of-sight halo}},
  \href{https://doi.org/10.1093/mnras/stac1967}{\emph{Mon. Not. R. Astron.
  Soc.} {\bfseries 515} (2022) 4391}
  [\href{https://arxiv.org/abs/2112.00749}{{\ttfamily 2112.00749}}].

\bibitem{inoue2023}
K.T.~{Inoue}, T.~{Minezaki}, S.~{Matsushita} and K.~{Nakanishi}, \emph{{ALMA
  Measurement of 10 kpc Scale Lensing-power Spectra toward the Lensed Quasar MG
  J0414+0534}}, \href{https://doi.org/10.3847/1538-4357/aceb5f}{\emph{\apj}
  {\bfseries 954} (2023) 197}
  [\href{https://arxiv.org/abs/2109.01168}{{\ttfamily 2109.01168}}].

\bibitem{inoue2005b}
K.T.~{Inoue} and M.~{Chiba}, \emph{{Three-dimensional Mapping of CDM
  Substructure at Submillimeter Wavelengths}},
  \href{https://doi.org/10.1086/452623}{\emph{\apj} {\bfseries 633} (2005) 23}
  [\href{https://arxiv.org/abs/astro-ph/0503212}{{\ttfamily
  astro-ph/0503212}}].

\bibitem{inoue2013}
K.T.~{Inoue}, V.~{Rashkov}, J.~{Silk} and P.~{Madau}, \emph{{Direct
  gravitational imaging of intermediate mass black holes in extragalactic
  haloes}}, \href{https://doi.org/10.1093/mnras/stt1425}{\emph{Mon. Not. R.
  Astron. Soc.} {\bfseries 435} (2013) 2092}
  [\href{https://arxiv.org/abs/1301.5067}{{\ttfamily 1301.5067}}].

\bibitem{erdl1993}
H.~{Erdl} and P.~{Schneider}, \emph{{Classification of the multiple deflection
  two point-mass gravitational lens models and application of catastrophe
  theory in lensing}}, {\emph{Astron. \& Astrophys.} {\bfseries 268} (1993)
  453}.

\bibitem{rennan1996}
R.~{Bar-Kana}, \emph{{Effect of Large-Scale Structure on Multiply Imaged
  Sources}}, \href{https://doi.org/10.1086/177666}{\emph{\apj} {\bfseries 468}
  (1996) 17} [\href{https://arxiv.org/abs/astro-ph/9511056}{{\ttfamily
  astro-ph/9511056}}].

\bibitem{mccully2014}
C.~{McCully}, C.R.~{Keeton}, K.C.~{Wong} and A.I.~{Zabludoff}, \emph{{A new
  hybrid framework to efficiently model lines of sight to gravitational
  lenses}}, \href{https://doi.org/10.1093/mnras/stu1316}{\emph{Mon. Not. R.
  Astron. Soc.} {\bfseries 443} (2014) 3631}
  [\href{https://arxiv.org/abs/1401.0197}{{\ttfamily 1401.0197}}].

\bibitem{mccully2017}
C.~{McCully}, C.R.~{Keeton}, K.C.~{Wong} and A.I.~{Zabludoff},
  \emph{{Quantifying Environmental and Line-of-sight Effects in Models of
  Strong Gravitational Lens Systems}},
  \href{https://doi.org/10.3847/1538-4357/836/1/141}{\emph{\apj} {\bfseries
  836} (2017) 141} [\href{https://arxiv.org/abs/1601.05417}{{\ttfamily
  1601.05417}}].

\bibitem{birrer2017}
S.~{Birrer}, C.~{Welschen}, A.~{Amara} and A.~{Refregier}, \emph{{Line-of-sight
  effects in strong lensing: putting theory into practice}},
  \href{https://doi.org/10.1088/1475-7516/2017/04/049}{\emph{Journal of
  Cosmology and Astroparticle Physics} {\bfseries 2017} (2017) 049}
  [\href{https://arxiv.org/abs/1610.01599}{{\ttfamily 1610.01599}}].

\bibitem{fleury2021}
P.~{Fleury}, J.~{Larena} and J.-P.~{Uzan}, \emph{{Line-of-sight effects in
  strong gravitational lensing}}, {\emph{arXiv e-prints} (2021)
  arXiv:2104.08883} [\href{https://arxiv.org/abs/2104.08883}{{\ttfamily
  2104.08883}}].

\bibitem{schneider2014}
P.~{Schneider}, \emph{{Can one determine cosmological parameters from
  multi-plane strong lens systems?}},
  \href{https://doi.org/10.1051/0004-6361/201424450}{\emph{Astron. \&
  Astrophys.} {\bfseries 568} (2014) L2}
  [\href{https://arxiv.org/abs/1406.6152}{{\ttfamily 1406.6152}}].

\bibitem{schneider2019}
P.~{Schneider}, \emph{{Generalized multi-plane gravitational lensing: time
  delays, recursive lens equation, and the mass-sheet transformation}},
  \href{https://doi.org/10.1051/0004-6361/201424881}{\emph{Astron. \&
  Astrophys.} {\bfseries 624} (2019) A54}.

\bibitem{zaldarriaga1998}
M.~{Zaldarriaga} and U.~{Seljak}, \emph{{Gravitational lensing effect on cosmic
  microwave background polarization}},
  \href{https://doi.org/10.1103/PhysRevD.58.023003}{\emph{\prd} {\bfseries 58}
  (1998) 023003} [\href{https://arxiv.org/abs/astro-ph/9803150}{{\ttfamily
  astro-ph/9803150}}].

\bibitem{lewis2006}
A.~{Lewis} and A.~{Challinor}, \emph{{Weak gravitational lensing of the CMB}},
  \href{https://doi.org/10.1016/j.physrep.2006.03.002}{\emph{Phys. Rep.}
  {\bfseries 429} (2006) 1}
  [\href{https://arxiv.org/abs/astro-ph/0601594}{{\ttfamily
  astro-ph/0601594}}].

\bibitem{ngVLA2018}
R.J.~{Selina}, E.J.~{Murphy}, M.~{McKinnon}, A.~{Beasley}, B.~{Butler},
  C.~{Carilli} et~al., \emph{{The ngVLA Reference Design}},  in \emph{Science
  with a Next Generation Very Large Array}, E.~{Murphy}, ed., vol.~517 of
  \emph{Astronomical Society of the Pacific Conference Series}, p.~15, Dec.,
  2018, \href{https://doi.org/10.48550/arXiv.1810.08197}{DOI}
  [\href{https://arxiv.org/abs/1810.08197}{{\ttfamily 1810.08197}}].

\end{thebibliography}\endgroup
\appendix
\section{Analytic Solutions for convergence and shear perturbations}
In the following, we present analytic solutions for approximated 'bare' convergence perturbation $\delta \kappa$ and shear perturbations $\delta \gamma_1, \delta \gamma_2$ for a foreground perturber, which are valid for $\beta \ll 1$. We assume that the magnification matrix of the dominant lens can be lineally approximated by three constants $c_1, c_2,$ and $c_3$ at $\x$ as described in eq. (\ref{eq:c1c2c3}). Plugging an approximated solution $\hat{\beta}^\textrm{A}$ in eq. (\ref{eq:beta-solution1}) into eq. (\ref{eq:ABCD2}), we have
\BEA
\delta \kappa
&=&-K^{-1}[2(-1+\hat{\beta}^\textrm{A} \kappa)(\hat{\beta}^\textrm{A}\gamma_2(B+\hat{\beta}^\textrm{A}c_3)-(A+\hat{\beta}^\textrm{A} c_1)
\nonumber
\\
&\times& (-1+\hat{\beta}^\textrm{A}(\gamma_1+\kappa))) -(A+\hat{\beta}^\textrm{A}c_1 -\hat{\beta}^\textrm{A}c_2 - D)
\nonumber
\\
&\times& (-(\hat{\beta}^\textrm{A})^2 \gamma_2^2 -(-1 + \hat{\beta}^\textrm{A}
 (\gamma_1 + \kappa))^2)],
\nonumber
\\
K 
&\equiv& 2 (-1 + \hat{\beta}^\textrm{A} (\gamma_1 + \kappa)) 
(-1 + 2 \hat{\beta}^\textrm{A} \kappa+(\hat{\beta}^\textrm{A})^2 (\gamma_1^2 + \gamma_2^2 - \kappa^2)),
\nonumber
\\
\EEA
\BEA
\delta \gamma_1
&=& L^{-1}(2 (B + \hat{\beta}^\textrm{A}  c_3) \gamma_1 - (A \hat{\beta}^\textrm{A} c_1 - \hat{\beta}^\textrm{A} c_2 - D) \gamma_2)
\nonumber
\\
&+&
M^{-1}[2 \gamma_1 (-((B + \hat{\beta}^\textrm{A} c_3) (1 + \hat{\beta}^\textrm{A} \gamma_1)) - 
   \hat{\beta}^\textrm{A} \gamma_2(\hat{\beta}^\textrm{A} c_2 + D) 
\nonumber
\\
&+& \hat{\beta}^\textrm{A} \kappa (B + \hat{\beta}^\textrm{A} c_3))],
\nonumber
\\
L&\equiv& 2 \gamma_2(-1+ \hat{\beta}^\textrm{A}(\gamma_1+\kappa)),
\nonumber
\\
M&\equiv& 2 \gamma_2(-1+2 \hat{\beta}^\textrm{A} \kappa+(\hat{\beta}^\textrm{A})^2(\gamma_1^2+\gamma_2^2-\kappa^2)),
\EEA
and
\BEA
\delta \gamma_2
&=& N^{-1} [B + \hat{\beta}^\textrm{A} (c_3 + (\hat{\beta}^\textrm{A} c_2 + D)\gamma_2 + \hat{\beta}^\textrm{A} c_3 (\gamma_1 
- \kappa)) 
\nonumber
\\
&+& B \hat{\beta}^\textrm{A} (\gamma_1 - \kappa)],
\nonumber
\\
N&\equiv& 1 - \hat{\beta}^\textrm{A} (2 \kappa + \hat{\beta}^\textrm{A} (\gamma_1^2 + \gamma_2^2 - \kappa^2)).
\EEA

%\bibliographystyle{mnras}
%\bibliography{2023BmodeLOS}

\end{document}